\documentclass[nolineno, apajournal,article,submit,moreauthors, fontset=macnew]{Definitions/mdpi} 
\makeatletter
\let\linenumbers\relax

\let\linelabel\@gobble
\def\@linenobox#1{}
\def\@linenoboxleft#1{}
\def\@linenoboxright#1{}
\makeatother
\usepackage{tabularx}    
\usepackage{makecell}    
\usepackage{booktabs}    
\usepackage{float}

\firstpage{1}

\pubvolume{10}
\issuenum{1}
\articlenumber{21}
\pubyear{2026}
\copyrightyear{2026}
\externaleditor{Martin Pohl}
\datereceived{3 January 2026}
\daterevised{26 May 2026}
\dateaccepted{27 May 2026}
\datepublished{3 June 2026}

\Title{Research on the Flat Field Measurement Method of Coronagraph}

\Author{Yulong Feng$^\ddagger$ 
 $^{1}$, Xuefei Zhang $^\ddagger$ * $^{2}$, Hongfei Liang $^{1}$*, Yu Liu $^{3}$, Mingzhe Sun $^{4}$, Tengfei Song $^{2}$ and Mingyu Zhao $^{2}$}

\address{%
$^{1}$ \quad School of Physics and Electronic Information, Chenggong Campus, Yunnan Normal University, Kunming 650504, Yunnan, China\\
$^{2}$ \quad Yunnan Observatories, Chinese Academy of Sciences, Kunming 650011, Yunnan, China\\
$^{3}$ \quad School of Physical Science and Technology, Southwest Jiaotong University, gdu 610068, Sichuan, China\\
$^{4}$ \quad Shandong University (Weihai Campus), Weihai 264209, Shandong, China}

\corres{Correspondence: zhangxuefei@ynao.ac.cn (X.Z.);  lianghongfei@ynnu.edu.cn(H.L.); Tel.: +86-13354610813 (H.L.)}

\secondnote{These authors contributed equally to this work.}

\abstract{The solar corona has an extremely low density, and its brightness is only about one millionth of that of the photosphere. High-dynamic-range imaging of its faint structure is therefore essential for studying coronal heating, coronal mass ejections, and space weather. Quantitative coronagraph imaging requires flat-field measurement and calibration, which underpin intensity calibration, small-scale feature detection, and long-term cyclic analysis. This paper analyzes the coronagraph imaging chain (baffle–optical system–detector) and the origins of flat-field errors, including optical aberrations, stray light, and pixel-response non-uniformity, and summarizes the resulting calibration requirements of next-generation coronagraphs. On this basis, ground-based and space-based flat-fielding methods are systematically reviewed: the ground-based methods include integrating-sphere uniform light sources, opal glass/diffuser plates, clear-sky and thin-cloud backgrounds, and solar-disk scanning, while the space-based methods include internal light sources and diffuser plates, attitude-roll and off-corona offset observations, and multi-phase statistical self-consistent flat-fielding. Their accuracy, resource cost, and applicability are compared. The review shows that no single method is simultaneously high-precision, easy to update, and engineer-friendly; a hierarchical, multi-method calibration framework is therefore recommended. Finally, a new method is proposed in which lithographically generated structured light fields, combined with Fourier-optics and machine-learning inversion, are used to estimate the pixel-response function. Preliminary experiments show that this method achieves a lower residual error than the integrating-sphere and opal-glass methods, providing a high-precision reference for future wide-band, high-resolution coronagraph calibration.}

\begin{document}

\captionsetup[figure]{name=Figure}
\captionsetup[table]{name=Table}

\setcounter{section}{0} 

\section{Introduction}

The corona is the outer region of the Sun's atmosphere, which has very low plasma density and extremely high temperatures of several million degrees; it extends far beyond the surface of the Sun.\cite{ref-1,ref-2,ref-3}. Under the above imaging conditions, in addition to noise caused by the detector charge-transfer process, flat-field errors are also among the main reasons for the reduction in detection sensitivity of faint coronal structures.\cite{ref-4,ref-5}.

The three parts of a typical occulting coronagraph are as follows: occulter, imaging system and detector. Theoretically, the flat field can be defined as the detector and optical-path response to spatially uniform incident illumination. Calibration and measurement of the flat field is also aimed at gaining more information about this response because the distribution of light brightness of a corona image should represent astronomical data instead of the property of some specific device\cite{ref-2,ref-6,ref-7}. A residual flat-field calibration error of even 1\% in the data products is sufficient to produce spurious signals comparable in brightness to weak coronal structures, severely hindering the identification and characterization of faint coronal features.

In recent years, with the deployment of active space coronagraphs that need to be self-consistent, flat-field calibration on-orbit has gradually become more necessary. The system has internal flat field lamps and diffuser plates which incorporate attitude roll data, and uses multi-point statistics to smooth out parameters using constant online scientific measurements~\cite{ref-8}.

New achievements in the field of nanostructures and lithography have provided a new path for fabricating controlled and spatially-structured light fields with rich information.~\cite{ref-9,ref-10}. This type of micro-/nano-scale mask with a train of flat-field coronagraph calibrations will enhance the spatial sampling resolution which is high in spatial frequencies and can be measured in one exposure. Fourier Optics and machine learning algorithms allow them to transform pixel responses into sub-pixel responses and overcome the constraints that may be imposed on accuracy, dynamic range and applicability of conventional flat-field methods~\cite{ref-11,ref-12}.

In this paper, motivated by the rapid development of micro-/nano-lithography technology, we propose and analyze a flat-field measurement method based on micro-/nano-patterned structured light fields for coronagraphs. It offers theoretical examples, and the concepts of design of experimental systems that promise to be a new promising way to do field calibration of the high precision and broadband coronagraphs in the future.

\section{Mechanism of Flat Field Error in Coronagraphs and Analysis of Measurement Requirements}

\subsection{Mechanism of Image Distortion in the Coronagraph Imaging Chain}

Three main elements can be identified in the imaging chain of the coronagraph: the occulter, the imaging system (optical system), and the detector~\cite{ref-2,ref-3,ref-13}. The occulter blocks the solar disk so that only the coronal radiation is permitted to pass through the subsequent optical path. This is a three stage chain wherein it is possible to see the effect of the baffle and the front end optics on stray light and low frequency background distribution and the overall effect of vignetting. Inhomogeneous detector pixel response contributes primarily to high spatial frequency non-uniformity in the image plane, as pixel-to-pixel response variations manifest as fine-scale structured patterns across the detector array~\cite{ref-2,ref-6,ref-14,ref-15}.

In such a hypothetical situation where one can assume that the incident light is almost uniformly spread across the view plane, the optical system does not have any aberrations, and detector pixels also behave in the same way then the brightness of the imaging plane ought to be uniform and it would achieve the ideal state of a flat field~\cite{ref-2,ref-13,ref-14}.
In terms of engineering, a residual flat-field error of 1 percent represents a change in brightness that is comparable to the measured contrast of an average K-class corona (2-3 solar radii). The signal-to-residual-error ratio is severely reduced by it in areas where there are faint coronal features-the effects of flat-field calibration errors are systematic and carry over between multiple science frames, so they cannot be eliminated by repeated measurements, causing unavoidable systematic bias in the measured coronal brightness distribution.

A coronagraph is an imaging device that is linearly spatially invariant or quasi-linearly spatially invariant as defined by systems theory. It is necessary to acquire a flat-field measurement, and thus the calibration needed should be proportional to the level of stability of the response of the system at the lower end of the spatial frequency range. It indicates that the output of every pixel is the same in case of identical incident light fields. The existence of spatially varying changes in point spread function including edge of field PSF expansion and higher scattering causes important contrast differences in luminescence~\cite{ref-16}. If the pixel response of the detector is not uniform across the pixels, then fixed-pattern noise will be added to the high-frequency portion.

High quality flat field calibration accuracy is not merely one instance of capturing an evenly illuminated image and normalizing it. This involves learning different types of errors including: optical aberrations, scattering, and response variation in detector pixels, which all can be described by the point spread function (PSF) of the system, or alternatively, its Fourier-domain representation, the modulation transfer function (MTF). This also accounts for recent researches that are using combined structured illumination fields with PSF models and deconvolution methods to get a physical inverse of flat-field~\cite{ref-12}
\subsection{Requirements of Flat-field Calibrations Accuracy and Stability}

The classical method of carrying out the flattening field calibration of using multi-pixel elements has failed to satisfy the requirements of quantitative study because the latest generation of space- and ground-based coronagraphs has significantly superior spatial and temporal resolution. The filament structure widths in high-resolution photos of a corona may be as little as several pixels wide and the variation in intensity at the start of a CME may be less than one-hundredth of a percent~\cite{ref-7}. This paper inspired the investigation of more sophisticated approaches like highly structured beams of light and extending the range of flat-field measurements to various homogeneous fields through scanning of fields of view to attain multi-mode and multi-spatial-frequency excitation with enhanced spatial resolution in inversion and so on.

Other types of coronagraphs can be used to measure all the plasma parameters at various temperatures and heights at multiple wavelengths (e.g., visible, near-infrared, and extreme ultraviolet)~\cite{ref-17}. Flat-field responses are extremely sensitive to wavelength: The same pixels may have entirely opposite responses at different wavelengths; the optical characteristics of vignetting and scattering are also wavelength-related.

To ensure this, it requires a method of flat field measurement that can offer consistent and comparable calibration across various wavelength bands, rather than being limited to a single operating wavelength~\cite{ref-18}. Coronagraphs have an intrinsic non-stationary response regardless of whether they are based on the ground or in space. Changes in the system on timescales of weeks to years may be due to contamination and aging of optical elements, degradation of coatings, detector quantum efficiency, dead pixels, and mechanical distortions because of thermal control~\cite{ref-19}.

The coronagraphs should be regularly updated through dynamic flat-field calibration to keep them up to date during their lifespan. Conclusively, the sources of the flat field errors are many; they are due to low frequency optical response and high frequency pixel non-uniformity, and also spatial non-uniform distribution between the pixels and the point spread function. Having this in mind, multi-scale or multi-modal excitation processes should be considered when developing future designs of flat-field mechanisms.

\subsection{Constraints for Flat-Field Measurement of the Coronagraph}

The coronagraphs based on the ground are typically placed in remote areas at high altitudes and in clean air without clouds or dust particles ~\cite{ref-7,ref-17}. Nevertheless, after a telescope is mounted on its structure, it becomes impractical and excessively expensive to return the system to the laboratory to calibrate it fully, so in situ calibration procedures are needed to preserve the flat-field precision during the working life of the equipmentt~\cite{ref-6}.

A few simple and user-friendly ways can be used to conduct flat-field measurements, such as frosted-glass or diffuser plates, a natural sky background, and thin clouds. The space coronagraph has been proposed to reduce the size, mass and power consumption of the satellite and enable in-orbit repairs~\cite{ref-8}. Therefore, the internal calibration systems need to be small, based on flat-field lamps and diffusers, and have mechanical limitations.

Additionally, the calibration process that uses satellite roll, or offset aiming, as part of the self-consistent flat-field method is also constrained because it will rely on how well the spacecraft attitude control is performed, the available attitude control resources (which may include reaction wheels or thruster fuel, depending on the spacecraft design), and how it impacts other payloads of the spacecraft under consideration~\cite{ref-20}. It should be noted that there are a number of other existing solar coronagraphs that have been shown to be effectively calibrated in flight with flat fields and do not require much attitude control resources. Examples include STEREO/SECCHI and METIS  that use stellar field drift scans at solar pointing to conduct flat-field calibration. An even more extreme case would be the PUNCH mission, which uses background star measurements and a steady spacecraft rotation in order to continuously achieve flat-field calibration during nominal science operation. To obtain the highest amount of calibration data at the lowest cost, it is required to implement orbital flat-field methods using the existing available operational science measurements.

The flat-field measurements of coronagraphs are also subject to trade-offs between accuracy, complexity, and resource consumption~\cite{ref-13}. The newest articles discussed the application of both classical approaches of the uniform light source and modern approaches such as statistical self-consistent inversion and structured light fields that rely on nanoscale lithography to achieve greater precision and capability at acceptable engineering costs~\cite{ref-6,ref-11}.

\section{Flat-field calibration Method for Ground-Based Coronagraphs}

\subsection{Integrating Sphere Uniform Light Source Method}

A laboratory approach has been taken to flat-field calibration using an integrating sphere uniform light source technique, providing near-ideal uniform-illumination conditions for full-aperture calibration of coronagraphs ~\cite{ref-2,ref-6,ref-13}. As one of the first and most widely used methods for flat-field calibration, it has good spatial uniformity and controllability in factory tests of ground-based and space-based coronagraphs. such as SOHO/LASCO~\cite{ref-21} and WISPR~\cite{ref-22}.

\begin{figure}[H]

\includegraphics[width=\textwidth, keepaspectratio]{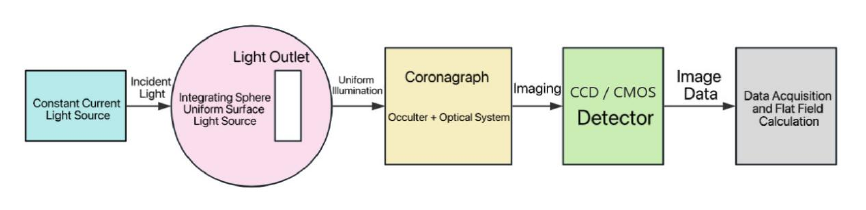}
\caption{Schematic of the optical path for flat-field calibration of the coronagraph with an integrating-sphere uniform light source.\label{fig1}}
\end{figure}

In case of conventional setup the output of the integrating sphere is connected to the optical portion of the coronagraph using either an imaging system or a collimator due to the fact that such approach will allow to characterize the pixel detector response (PRNU) non-uniformity and low-frequency optical vignetting simultaneously in highly controlled conditions, see Figure~\ref{fig1}.

\unskip

\begin{figure}[H]

\includegraphics[width=\textwidth, keepaspectratio]{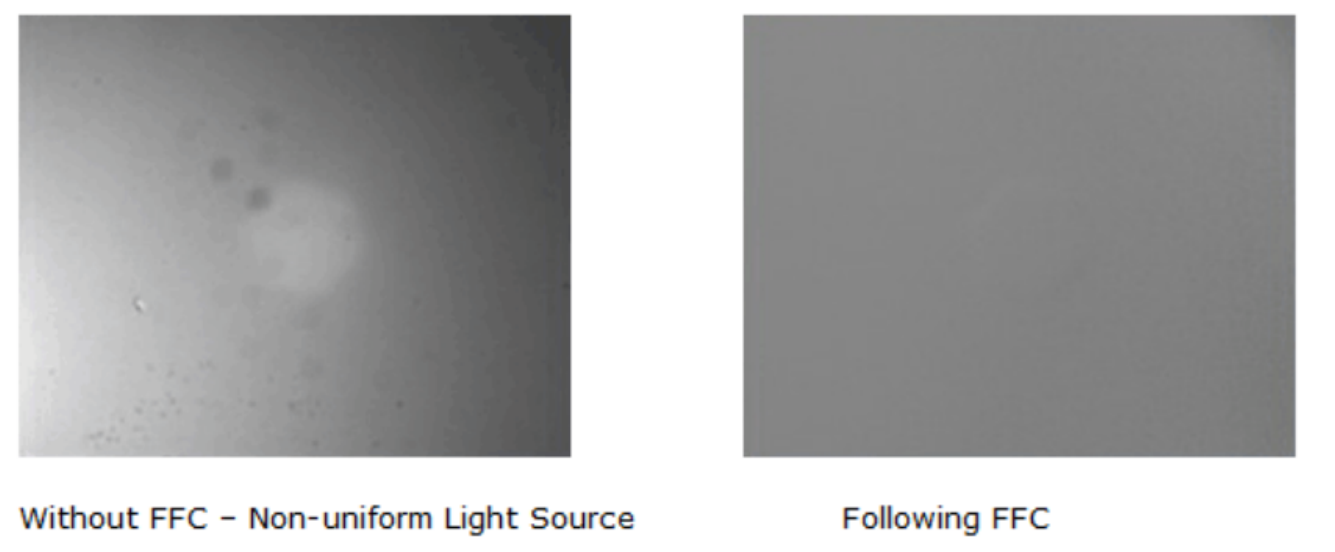}
\caption{Residual distribution schematic before and after application of flat-field correction of integrating sphere or uniform light source~\cite{ref-14}. (Left): Residual map before correction, showing distinct radial background gradient; (Right): Residual map after flat-field correction, in which systematic structures are substantially reduced and remaining residuals are predominantly random noise.\label{fig2}}
\end{figure}

The root-mean-square error (RMSE) of the 2048×2048 detector at 550 nm is experimentally measured to be no more than 0.42±0.03 percent in the middle part of the image that has been measured in the form of an area consisting of 512x512 pixels. It can also be stated concerning SOHO/LASCO C2 in which RMSE in visible-light flat-field measurements was discovered to be within the range of 0.4\% to 0.7 percent~\cite{ref-2}. The true flat-field uncertainty is stable at about 0.4 percent and includes contributions from PRNU, dark current, and read noisee~\cite{ref-23}. The uniformity of light output of an integrating sphere system is reported as 99.0–99.5 percent, and the PRNU-corrected calibration uncertainty is about 0.2–0.5 percent in the visible and near-infrared regions. The listed properties enable the integrating sphere flat field to function as a very precise lab calibrated standard that may be used in subsequent quantitative measurements of corona brightness. As shown in Figure~\ref{fig2}~\cite{ref-24}, the residual distribution before and after using flat-field corrections shows that there are fewer large scale radial gradients and systematic structures.

\unskip

This method would be expensive in terms of hardware costs, its implementation would also be complicated, and any calibration would take 2 to 4 hours. Furthermore, it cannot reproduce the stray light, thermal and mechanical effects that actually occur on site during actual observation. Although the integrating sphere approach has the highest calibration accuracy, it is not practical for daily use; thus, a low-frequency baseline flat field must be provided instead of a tool for regular operational support under field observation conditions~\cite{ref-2,ref-7}.

\subsection{Flat-field method for opal glass/diffuser plate}

In-situ calibration techniques using flat-field devices with opal glass or diffuser plates are commonly applied to ground-based coronagraphs. It follows that this technique is based on introducing a high-Lambertian diffusion piece into the first point of the light pathway where it generates quasi constant illumination fields which can be transmitted through similar optical geometries as that which exists during scientific observation. Not necessarily ideal laboratory conditions but actual working conditions can be applied to do so.

It has also been shown that this method works well and with high efficiency. In more detail, Bai Xianyong et al. examined the process of implementation, the main causes of errors and the quality of calibration of the frosted method in the systematic fashion to conclude that this method was simple to set up and offered some degree of flexibility, it has great repeatability and is among the most often utilized in-situ flat-field solutions for ground-based coronagraphs~\cite{ref-6,ref-24}. The findings of the experiment indicate that the stated methodology is very efficient in reducing the low-frequency background perturbation and the detector response change and provides a fair tradeoff between the precision of the calibration and operational efficacy.

\begin{figure}[H]

\includegraphics[width=\textwidth, keepaspectratio]{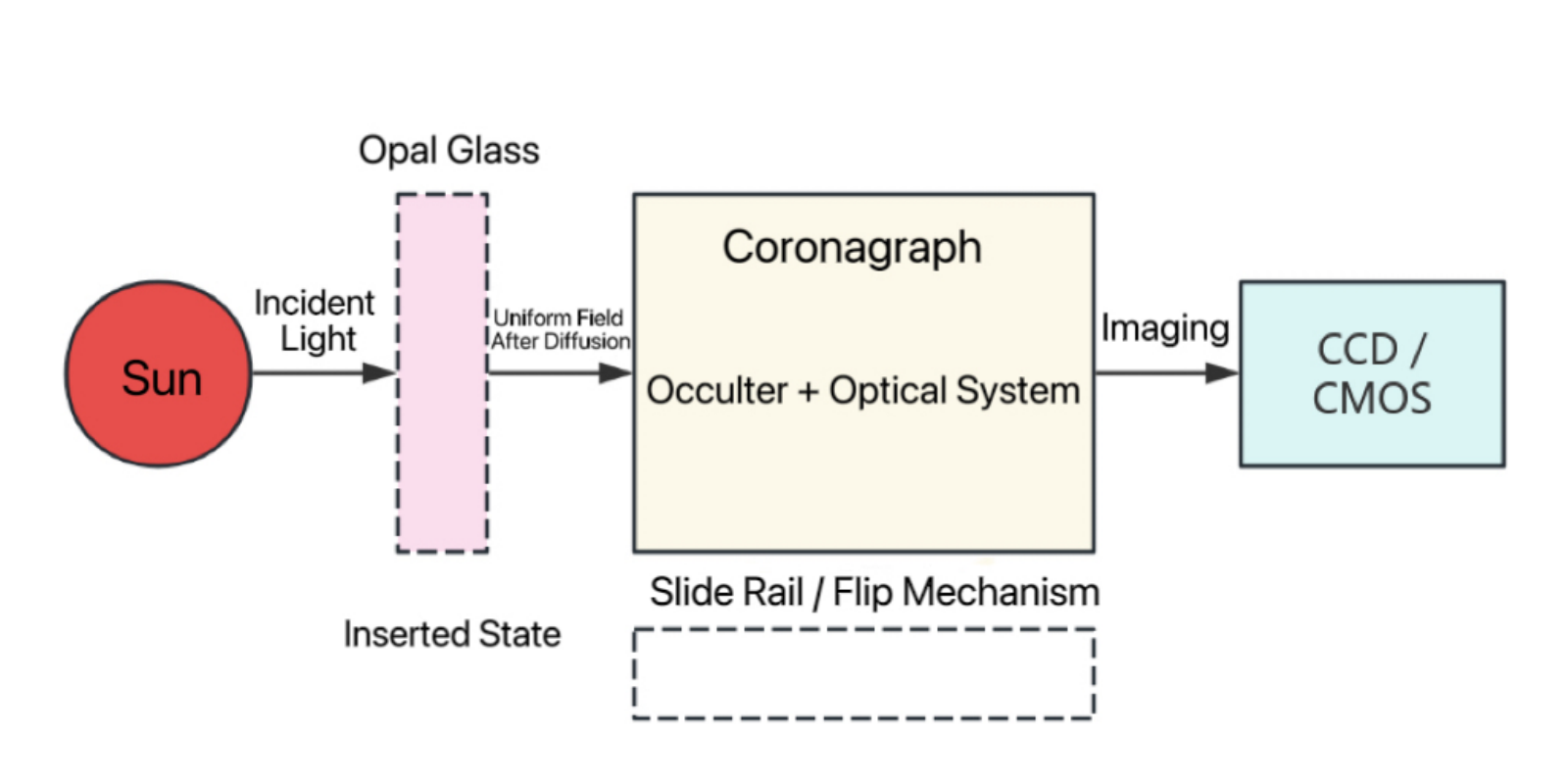}
\caption{Schematic diagram of flat-field calibration achieved by inserting an opal glass/diffuser plate into the front end of the coronagraph.\label{fig3}}
\end{figure}

The diffuser plate is usually placed at the entrance to the coronagraph in either of the two mechanisms (slide or flip) as shown in Figure~\ref{fig3}.

\unskip

Since the scattered light follows the same optical path as during regular observation, the resulting flat field captures effects such as gravity-induced flexure, thermal deformation, and optical contamination. According to the experience of engineers, various mechanical devices (including inserts and direct inserts) and flip-over systems have almost equal quality of correction and the choice depends largely on both the design of the device and ease of operating its mechanism.

\begin{figure}[H]

\includegraphics[width=\textwidth, keepaspectratio]{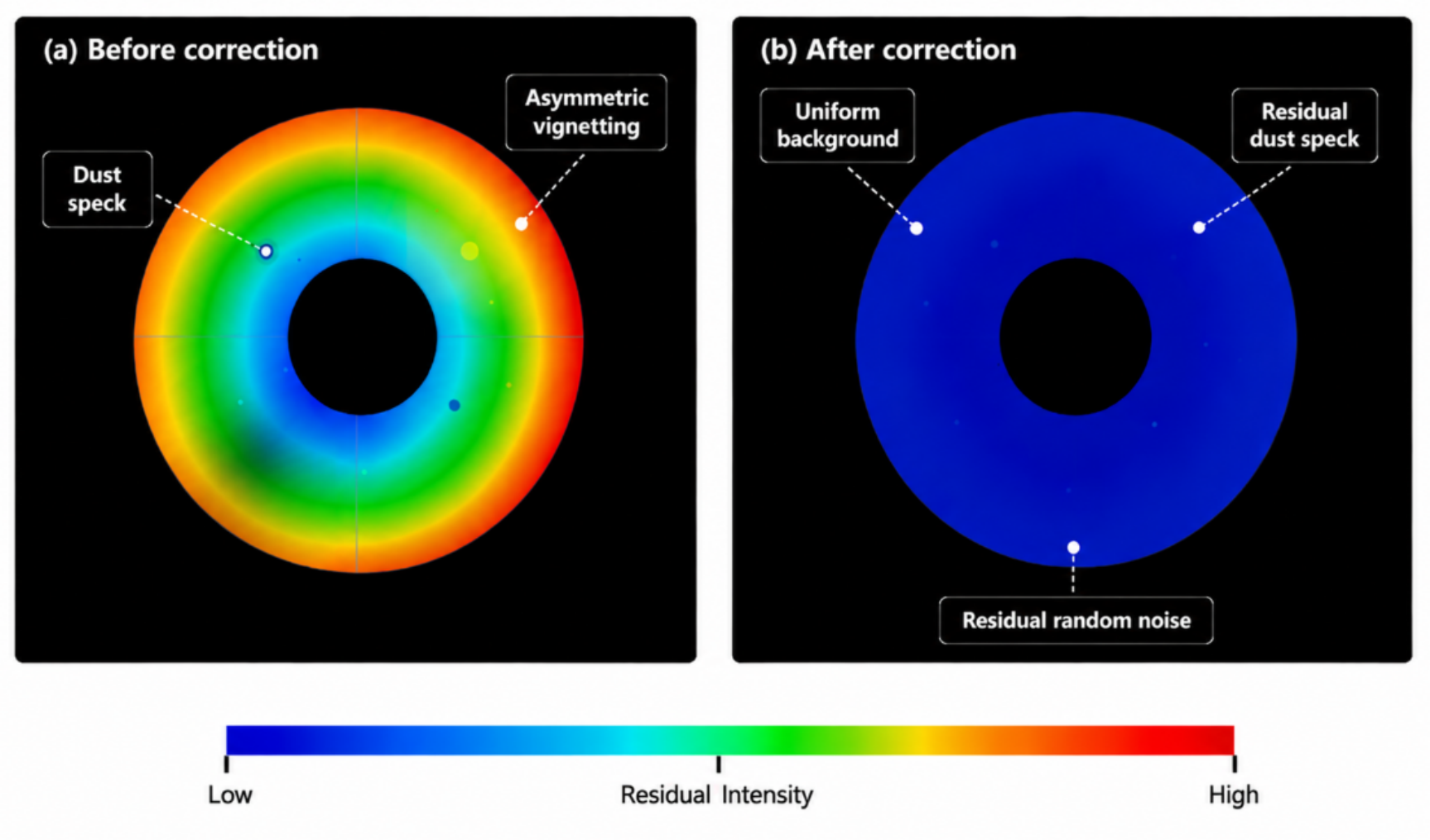}
\caption{Residual distribution maps before and after flat-field correction with frosted glass/diffuser plate. Left (a): Map of residual before the flat-field correction with a strong large-scale radial vignetting gradient shown in pseudo-color. Right (b): Residual image with frosted glass flat-field correction, which demonstrates a significant reduction in the systematic low-frequency background structure; the left-over residuals are primarily random noise.\label{fig4}}
\end{figure}

Quantitative and qualitative changes achieved by this method are represented in Figure~\ref{fig4}, which shows remaining distribution maps prior to and after diffuser-induced plane correction, shown in pseudo-color to emphasize systematic background structure.

\unskip

The residual maps show how the major part of the large-scale radial background gradients is removed in the course of correction and the rest of the residuals are mostly random noise instead of systematic low-frequency artifacts. Published data has indicated that in the best conditions, the residual readings of the flat-field RMSE will reach the 1 percent range~\cite{ref-6}. Similarly measured under visible light, compared with the absolute level of the integrating sphere approach (RMSE = 0.42 ± 0.03), this approach has a slightly larger overall RMSE of 0.85 ± 0.08\% on average. The diffuser plate technology is less accurate but may be used to deliver a constant output with repeatability and can be applied to normal field calibration. Therefore, it has been indicated that the given strategy cannot be used to get the final flat-field reference values, but it is extremely useful to monitor and remedy the tendency of low frequency in flat fields when operating in the ground with a long duration.

\subsection{Natural Sky Background/Thin Cloud Horizon Method}

Thinner cloud flat-field and natural-sky background approach is based on the fact that the sunlight scattered both by the air molecules and by the aerosols results in Mie-Rayleigh scattering. Geometrical and atmospheric conditions are so that can estimate this system and the sky atmosphere as the large-scale, quasiuniform illumination source and be one of the examples of what is known under the name of the natural integration sphere in calibrating ground-based coronagraphs with the help of a flat-field method~\cite{ref-2,ref-14}.

Observationally it is proven that in case of altitude of the site being more than 3000 meters even in the case of good weather conditions, the sky background does not change significantly on most occasions, and not only during the full eclipses of the sun but also partial or annular ones. According to the CATEcor measurements of the annular eclipse of 14 October 2023, it has been represented that appropriate portions of the sky may serve as the indicators of low frequency flatfield. As shown in Figure 5, when the sun is at a low elevation (dawn/dust) or when the sky is partially covered by thin clouds (low-scattering), multi-scattering can erase the localized changes in intensity at angles larger than the coronagraph field of view.

\begin{figure}[H]

\includegraphics[width=\textwidth, keepaspectratio]{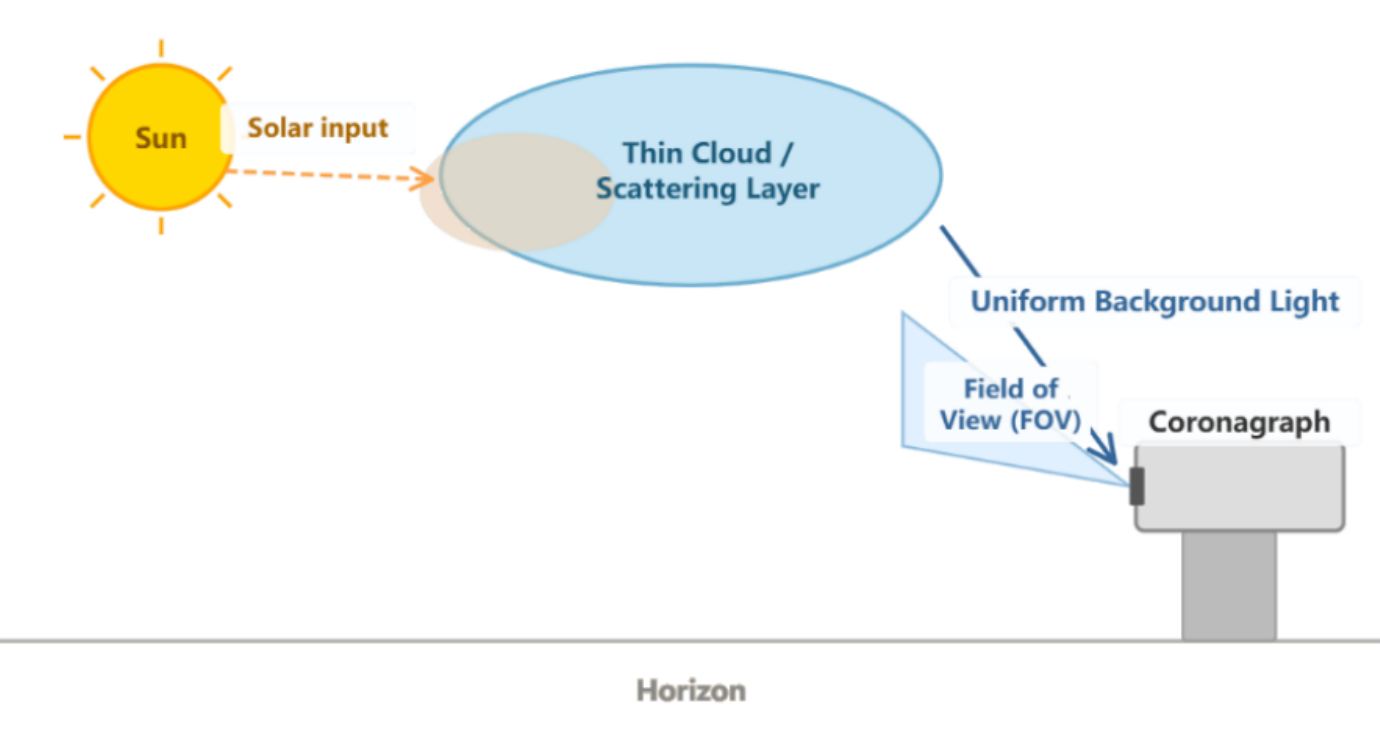}
\caption{Schematic diagram of the observation geometry for flat-field calibration using a natural-sky background/thin clouds.\label{fig5}}
\end{figure}
\unskip

The technique requires a stringent limitation in the formulation process of observation plans when contrasted with instrumentation arrangements. These measurements are usually made at some distance away from the direct view of the sun, and the measurements are taken in several brief intervals, multi-frame measurements so that they will span the entire dynamic range of the detector. The literature states that important causes of errors are the nonstationarity of the brightness of the sky with time, the scattering spectrum and polarization, and the interaction of clouds gradient with the instrument vignetting.When acquiring around 100 or 300 images in a typical 10-30 minute period, the RSME of low frequency residuals of flat backgrounds can be minimized to less than 0.8-1.5\%~\cite{ref-2,ref-14} provided that a reasonable choice of frames and background is made to achieve similar outcomes of this kind of workflow. With high frequency response factors of a pixel scale, the average RMSE tends to rise up to two or five percent, which is very sensitive to the weather.

\begin{figure}[H]

\includegraphics[width=\textwidth, keepaspectratio]{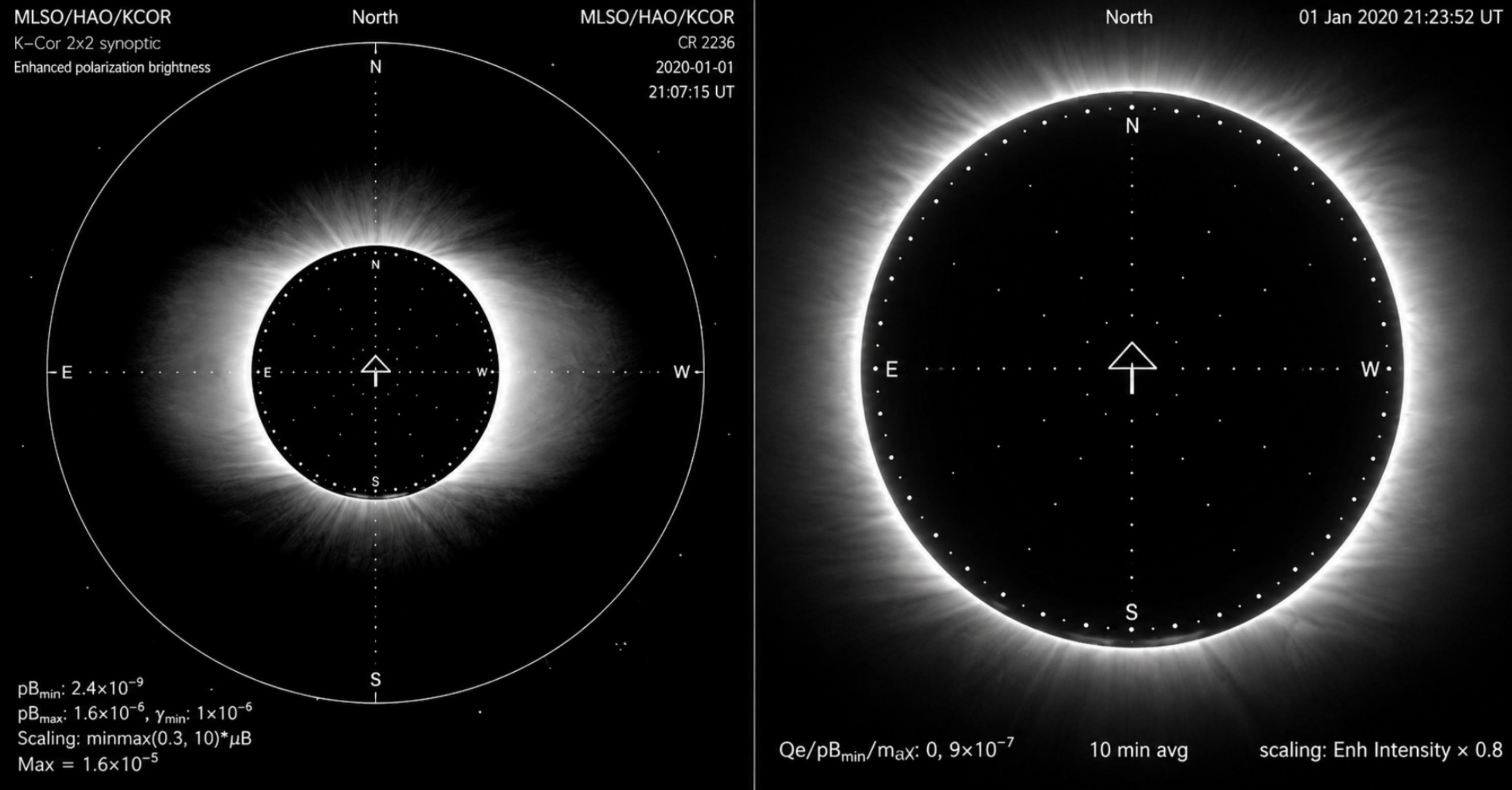}
\caption{Example of a real observation image from a ground-based white-light coronagraph under natural sky background conditions. The right panel is a magnified view of the left panel, shown at enhanced scale to highlight the coronal spatial structure and directional orientation (N/S/E/W). Intensity scaling: min/max = 0–9×10$^{-7}$, 10-minute average. (Data source: Mauna Loa Solar Observatory, K-Cor Level-2 Polarized Brightness (pB) Data, 2020-01-01, HAO/NCAR).\label{fig6}}
\end{figure}

Figure~\ref{fig6} presents some example observational data recorded in natural sky conditions which indicate that the output illumination field is nearly uniform as opposed to being perfectly uniform.

\unskip

(Data source: Mauna Loa Solar Observatory, K-Cor Level-2 Polarized Brightness (pB) Data, 2020-01-01, HAO/NCAR). This method may also prove useful in a flat field calibration since it controls large scale and lower frequency variations in background, but much smaller elements are present to correct pixel-to-pixel response variation due to different pixel environments. Although it is simple and easy to implement in hardware, it is very sensitive to atmospheric stability and length of monitoring. Therefore, thin-clouds/natural skies approach is not so much appropriate to monitor long-term trends because the sky background itself changes over the course of the observation. The method is more properly used to check approximate flat-field calibration in one-time measurements when detailed calibration procedures cannot be performed-the CATEcor observations that were taken during the annular solar eclipse of 14 October 2023 represent the best example of these measurements~\cite{ref-1,ref-2,ref-14}.

\subsection{Solar Disc Scanning and Sweeping Method}

The scanning of the solar disk and scanning techniques rely on a significant a priori hypothesis: Specifically, the spatial scales are large enough and adequate time averaging is considered so that the distribution of brightness of the solar photosphere may be considered smooth and may possess a parametric description~\cite{ref-25}. Under such an assumption, small movements across the solar disk will result in the fact that the identical pixels in the detector will see a different region of the solar disk with time.

In contrast to the uniform light source methods, this one does not employ an outside reference light source, instead, it uses the form of the heavenly body in space as its self-calibrator. Practically speaking, solar disk scan technique typically uses small-scale RA(Right Ascension) and Dec(Declination) control which lead to two-dimensional scanning lattice across the solar disk. It is a data cuboid where time, pointing information and brightness information are the dimensions. With the help of such a data structure it could be possible to construct an observational model in the following way:
\begin{equation}
\begin{split}
\mathit{Observed\ signal} = {} & \mathit{Solar\ reference\ brightness} \times \mathit{Pixel\ response} \\
& {} \times \mathit{Optical\ path\ transmission} + \mathit{Background\ scattering\ term}
\end{split}
\end{equation}

Inversion is performed in three stages. Firstly, a parametric limb-darkening polynomial model is adjusted to the solar disk brightness distribution at every particular scan pointing location, which serves as a Solar reference brightness term. Secondly, the Pixel response function and Background scattering term are estimated concurrently using a regularized least-squares optimization over all multi-point, multi-temporal measurements, assuming this solar reference. In particular, due to the time dependency of the pixel response and background scattering being slow compared to the time dependency of the solar brightness being dependent on the pointing geometry, the two parts may be effectively separated with adequate spatial and time sampling. Finally, the flat-field template is obtained using this combined fit as the pixel response portion, less the fitted background scattering term. Prior to fitting active regions and solar flares are masked and excluded to prevent contamination of the flat-field estimate with transient solar features.

\begin{figure}[H]

\includegraphics[width=\textwidth, keepaspectratio]{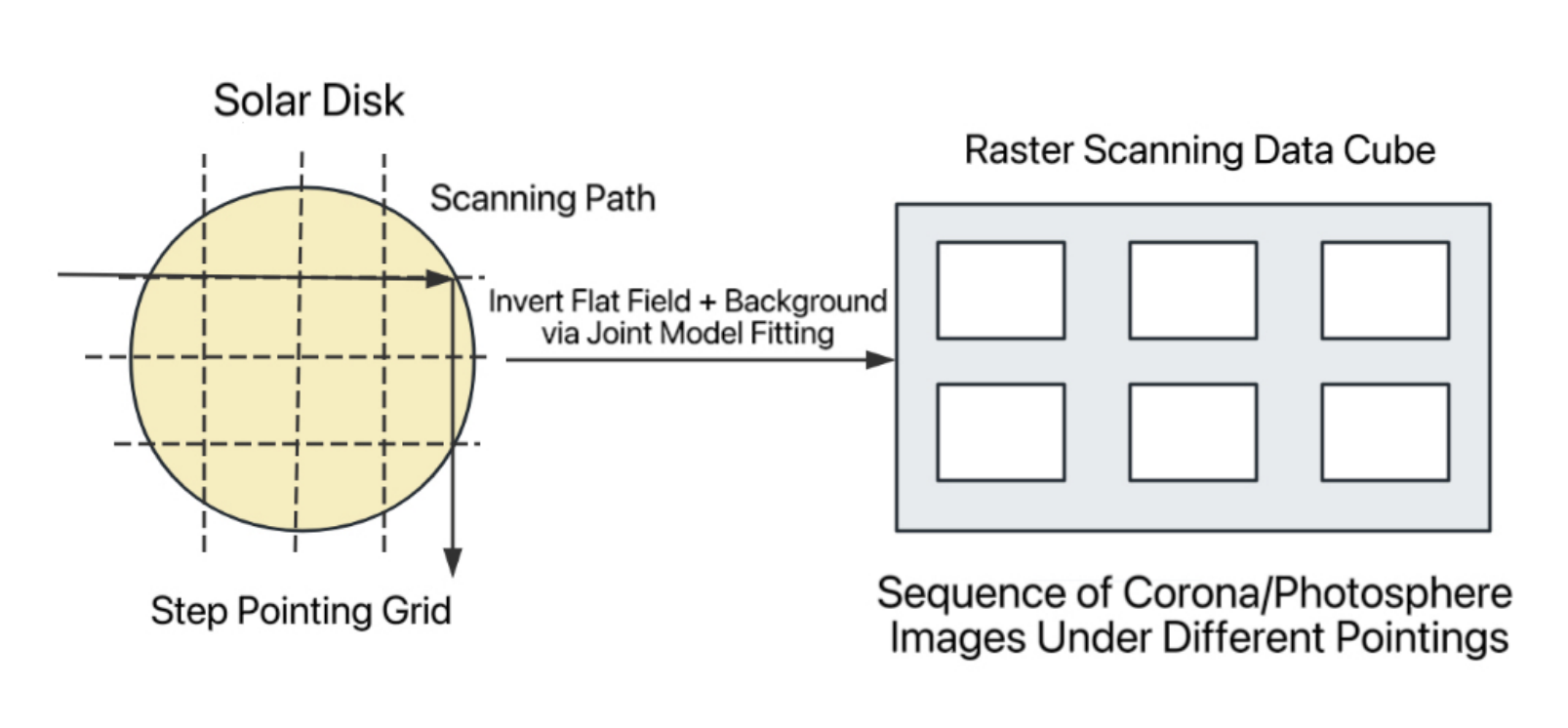}
\caption{Schematic diagram of the pointing grid and data cube for flat-field calibration of the solar disk scan (self-drawn diagram).\label{fig7}}
\end{figure}
\unskip

A measurement of the non-uniformity of pixels responses and scatter owing to the background at once in a uniform model is through multi-point, multi-temporal joint-fitting of observation data, which is similar to flat-field and background separation, as shown in Figure \ref{fig7}. The published engineering papers have shown that this joint inversion procedure has the potential to simultaneously estimate the pixel-scale response non-uniformity as well as the large-scale background scattering elements of the flat-field with the residual RMSE of about 1-2 percent under controlled observing conditions~\cite{ref-14}. The approach is extremely susceptible to errors of pointing and the complexity of the natural solar structure, as well as alterations in the meteorological conditions of seeing and scattering in the event of homogeneous quantization.Based on engineering practice, one may say that when the number of scanning points is not less than 8 8 and 8 frames are taken at every position of the grid then the average is computed across short sequences and all active areas are removed strictly, the resultant Root Mean Square Error (RMSE) with combined fit is usually in the vicinity of 0.8 - 1.8 per cent~\cite{ref-2,ref-5}. It has also been observed how poor calibration can be witnessed due to the presence of poor repeatability of guidance or the fact that solar activities have changed during the scanning process as evidenced by the residual values of RMSE that can be even up to 2–4\%.

\begin{figure}[H]

\includegraphics[width=\textwidth, keepaspectratio]{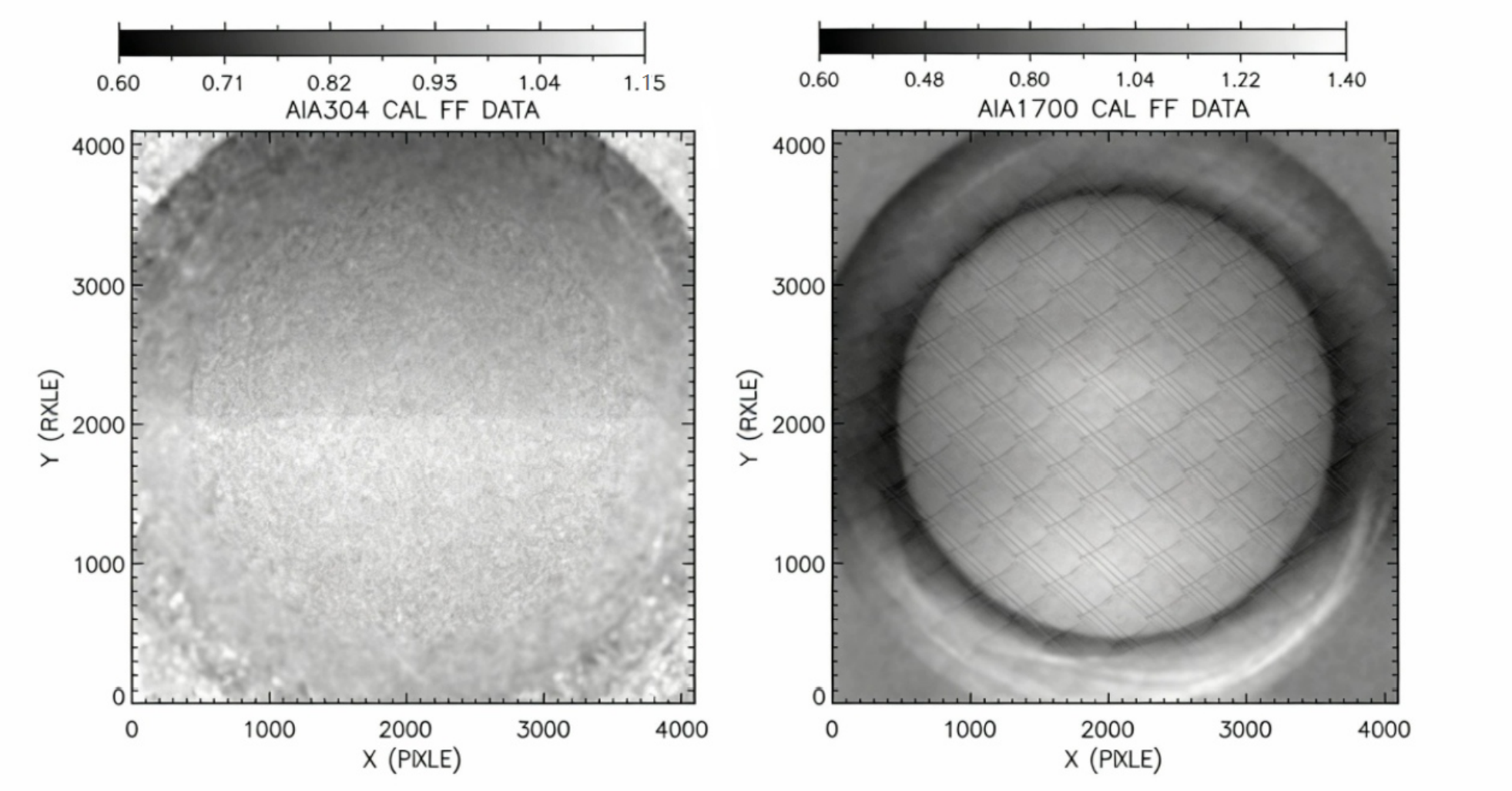}
\caption{Sample Solar Disk Scan Flat Field Template Depending on a Reconstructed Scan Inversion Algorithm (Data source: Solar Dynamics Observatory / Atmospheric Imaging Assembly, SDO/AIA).\label{fig8}}
\end{figure}

An 8×8 scan process typically takes 30–90 minutes, and may require longer depending on additional time costs. Then estimate dislocation, construct models and joint fitting. Multi-point and multi-temporal data are likely to produce specific information on solar disk and pixel responses by utilizing solar disk and field scanning techniques. A representative reconstructed flat-field template obtained from this scan inversion algorithm is shown in Figure~\ref{fig8}.
\unskip

From a quantitative point of view, published engineering studies show that with properly planned scanning procedures the size of the grid is at least 8×8 points, more than one frame per pointing position as well as the efficient suppression of active regions and periodic adjustment of pointing positions, it is possible to decrease residual RMSE (root mean square errors) by several percent up to 1 percent~\cite{ref-2,ref-5}. This new degree of spatial consistency makes the new data base able to become more homogeneous, which could be applied to supply supplementary background model building, modeling of weak features, and assessing changes over time.

On the contrary, instead of employing laboratory-based methods in which artificial uniform light sources are used during the calibration of flats, performance statistics, RMSE and spatial consistency indices are generally used to indicate the worth of flat-field correction in solar disk scanning in reality, not necessarily by evaluating one frame at a time but all together.

Astronomy and engineering are the cause of error in the solar disk-scanning and field-of-view-scanning technique. Changes in the behavior of sunspots, along with other events of lesser length like solar flares, may not be consistent with the original hypothesis of constant brightness during scanning~\cite{ref-1}. An engineer can assume that the errors caused by pointing are the reasons why there are discrepancies between actual scan location measurements and forecasted scans and consequently leads to systematic residuals~\cite{ref-7}, and it will also be coupled by the change in the seeing and scattering characteristics of the atmosphere that will result in a decrease in the stability of effective illumination field over time.

These and all other factors will determine a proper degree of solar disk scanning flat-field calibration (and should be paid special attention when comparing the different observational conditions and selecting between the models and data). On the whole, it can be concluded that the offered methodology may be successfully implemented in the case when the working conditions of solar telescopes are kept constant.

\subsection{The evaluation of the site flat-field calibration methods in detail}

The integration sphere uniform light source technique was also applied in the laboratory component to acquire highly uniform responses of pixel non-uniformity and low frequency gradients. It serves as the foundational flat-field reference for subsequent in-situ updates planned throughout the operational lifetime of the instrument. It is one of the in situ update lines with more in situ upgrades planned. The in-situ upgrades are most effective in practice at the station with use of the frosted glass/diffuser plate method and method of natural sky background. Another way of frosted glass might be of use to limit thermal-gravity drift and misalignment while still maintaining the right optical path geometry and alignment to gravity of the instrument. Thin cloud cover or the natural sky where the refresh rate of low-frequency flat-field data is fast with no extra equipment. Solar disk scans and field of view scans of older design can be added to uniform-light-source methods as geometrically self-calibration steps can be achieved when sufficient pointing stability and solar brightness can be predicted.

\begin{table}[H] 

\caption{Quantitative comparison of flat-field calibration methods for ground-based coronagraphs (under equal instrument conditions).\label{tab-comparison}}

\begin{tabularx}{\textwidth}{p{2.9cm} p{2.7cm} p{2.5cm} CC}
\toprule
\textbf{Method}	
& \textbf{\textbf{\parbox{2.5cm}{\centering Root Mean Square Error \\ (RMSE) (\%)}}}
& \textbf{Calibration time required}	
& \textbf{Hardware costs}	
& \textbf{Repeatability ($\sigma$)}\\
\midrule
Integrating Sphere Method for Uniform Light Sources	& 0.42$\pm$0.03~\cite{ref-2,ref-8}	& 3.1$\pm$0.2h~\cite{ref-2,ref-7}	& $\sim$\,USD\,48{,}000	& 0.04\%\\
Natural Sky Background/Thin Cloud Method	& 0.85$\pm$0.08~\cite{ref-6}	& 0.4$\pm$0.1h~\cite{ref-23}	& $\sim$\,USD\,2{,}800	& 0.08\%\\
Opal Glass/Diffuser Panel Method	& 2.7$\pm$0.3~\cite{ref-7,ref-14}	& 0.3$\pm$0.1h~\cite{ref-23}	& $\sim$\,USD\,0	& 0.32\%\\
Solar Disk Scanning/Field Sweeping Method	& 1.2$\pm$0.15~\cite{ref-4,ref-26}	& 1.0$\pm$0.2h~\cite{ref-2,ref-5}	& $\sim$\,USD\,11{,}000	& 0.12\%\\
\bottomrule
\end{tabularx}
\end{table}

Table~\ref{tab-comparison} shows that in the case where similar conditions of instruments and wavelengths are used, the integrating-sphere method is the most accurate technique for quantifying flat fields, but it consumes the most significant period of time and engineering capital. The frosted-glass method strikes a good balance between high accuracy and high efficiency, and it is an emerging solution for in situ updates of ground-based coronagraphs.

Based on the analysis of ground-based methods above, three main conclusions can be drawn: Firstly, the two-stage calibration system must be adopted, which is the combination of both high laboratory accuracy and updating the work onsite. Secondly, it is worth noting that standard illumination cannot necessarily eradicate stray light and in situ deviation; however, geometric self-consistency and statistical self-consistency supplement each other.

Additionally, there are flat field templates that need to maintain a trackable updated data chain. These findings can also be extended to the development of the on-orbit calibration system of space coronagraphs and the development of the new structured light field technique suggested in this paper.

\section{Flat-Field Calibration Method for Space Coronagraphs}

\subsection{In-Orbit Flat-Fielding Method for Internal Light Sources and Diffusion Plates}

It may be that this is a more advanced, technology-driven way among in-orbit flat-field methods, utilising onboard light sources and diffuser plates for space coronagraphs. Generally speaking, introduce a stable and repeatable artificial light source inside the instrument that is powered by a regulator. When a given light source and a particular diffuser design are used, the detector end of that device achieves almost uniform illuminated region.

At present, the most frequently used kinds of internal light sources are light-emitting diodes (LEDs) and halogen bulbs because of their stable emission properties and long-life characteristics. During the flat-field acquisition process, the solar pointing direction of the instrument should not change. Then the internal light source is turned on and the detector collects flat-field data in the same thermal and electronic environment as the scientific observations. Flat-field images which have been normalized and statistically averaged can result in a flat-field template representing the non-uniform response of the system~\cite{ref-18,ref-25}.

For this method, the following conditions are required: the internal source must maintain stable emission throughout the mission lifetime; the diffuser must not exhibit considerable temporal variation in scattering characteristics; and the alignment between the internal illumination and science observation optical paths at the detector must be sufficiently good~\cite{ref-22}.

The regular in-flight calibration with an internal light source and without changing the instruments position relative to the sun or its activity creates a flat-field template that is the difference between the pixels of the detectors and the optical paths. The provided flat-field template has served as one of the steps of the data processing to normalize raw observational data in order to reduce systematic background variation due to non-uniformity of pixel responses and variations in optical path~\cite{ref-2,ref-4,ref-5}.

Studies have shown that for the Metis coronagraph visible-light channel on the Solar Orbiter mission, residual flat-field non-uniformities can be stabilized to around 1\% by periodically activating the internal light source to acquire in-orbit flat-field templates~\cite{ref-27}. Assess the stability of the background to confirm that the system has achieved a reasonable level of stability; thus, inverting the polarised brightness (pB) and detecting weak coronal structures will be more reliable across different observation periods and more consistent in the data.

However, several disadvantages can be noted. The internal light source and diffuser belong to the instrument so that the illumination field is accompanied by some internal scattering pathway and thus it does not reproduce the stray light distribution that would be produced in actual coronagraph observations. In addition, the internal light sources can have a limited spectral range that is non-compatible with the scientific observation band and therefore cannot be used in wide-band or multi-channel systems. The pixel-to-pixel response nonuniformity of the detector is measured by an in-orbit flat-field calibration with an internal light source, which is usually applied as a supplement to the higher-precision baseline set during the ground-based flat-field calibration before launch, but not as its substitute.

\subsection{Attitude Roll and Solar Corona Offset Observation Method}

The roll maneuver field-of-view flat-field technique is a small-angle discrete roll or a full-cycle continuous roll around the satellite's optical axis, performed with a stable observation pointing and Sun-satellite distance. The assumption behind it is as follows: Continuously observed rolls show a relatively uniform distribution of coronal brightness and are statistically axisymmetric.

Under this assumption, regardless of the apparent motion of coronal structures across the detector field of view at different rotation angles, the brightness of the corona at the same radial distance and azimuth should be statistically constant. Multi-roll-angle /multi-temporal collaborative modelling may be applied to deconvolute the intrinsic distributions of coronal brightness into detector pixel response and optical path transmission variations to give a self-consistent inversion of the flat-field function.~\cite{ref-4}

Attitude roll is one of the most significant methods to conduct the in-orbit validation and cross-calibration, especially in the case of the Metis coronagraph on Solar Orbiter. The ratio between polarized intensity and total intensity (pB/I) measured at the same radial height but with different roll angles can be used as an indicator of the quantitative level of a residual polarization flat field and system response stability. In-orbit measurements show that with appropriate selection of coronal regions excluding active regions, variations of pB/I across roll angles can be controlled to within only one or two percent, which is regarded as an acceptable residual level for polarization flat-field calibration~\cite{ref-20,ref-28}.

Attitude maneuvering and observation time is also restricted and may cause the irregularities in the process of science observation. It presupposes statistical stationarity and spatial isotropy of the corona during the observation period. Nevertheless, significant coronal mass ejections, solar flare eruptions or highly active regions are capable of upsetting the assumptions made by the models~\cite{ref-1,ref-4,ref-18}.

\subsection{Multi-Phase Statistics-Based ``Self-Consistent Leveling'' Inversion Method}

Recent methods that combine self-consistency with multi-temporal statistics have gained popularity in space-based solar observation and coronagraph data processing as data-driven approaches to flat-field calibration. Hence, continual computation and reconstruction of the flat-field can be performed without additional on-orbit hardware~\cite{ref-2,ref-25}.

The fundamental idea behind this kind of approach is that it has a broader period. The characteristics of instrumental pixel response and optical path transmission often vary very slowly with time, whereas solar corona brightness profiles can be quickly changed depending on the time and the geometry of observation~\cite{ref-5}.

Mathematically, such methods typically represent the observational data at the $i$-th time step as
\begin{equation}
D_i(x,y) = F(x,y)\cdot I_i(x,y) + B(x,y) + N_i(x,y) ,
\end{equation}

Here $F(x,y)$ is the flat-field function to be solved, $I_i(x,y)$ is the coronal brightness distribution at the $i$-th time step, $B(x,y)$ is the background scattering term that is nearly time-invariant, and $N_i(x,y)$ is the noise term.

Engineering practice in space coronagraph data processing shows that: One self-consistent model of the flat-field inversion is probably to decrease the inhomogeneity of the detector response and eliminate specific low-frequency background features if the observational data has enough temporal and field-of-view variability. Long-term observations were made using space missions such as the Parker Solar Probe/WISPR and SOHO/LASCO with a constant set of normal selections of observations and controlled prior inversions which allowed the residual non-uniformity after flat-field inversion can generally be reduced to less than or equal to 1–2\%. Its size is constrained to this amount because of the inseparable link of large-scale light distribution in the corona and the flat field function.There are various restrictions to the application. The approach assumes a background hypothesis, which is that the flat-field response will not change significantly over the period of inversion time. When there is contaminant build-up in the instrument over a short timeframe, thermal deformations, or sudden changes in the electrical states of the instrument, the inversion results may be systematically biased. It is especially so when the active regions or recurrent coronal mass ejections are observed~\cite{ref-25}.

The self-consistent flat-field inversion algorithms can also be viewed as another form of flat-field calibration applied over longer time scales than short-term methods, used to assess in-orbit stability~\cite{ref-29}. They must not be considered as a separate group of calibration instruments with the ability to accurately measure the ground flat-field. In practice, it can be applied to develop more robust flat-field calibration methods in combination with other methods, such as the internal light source flat-fielding, attitude roll observation, or the offset observation method.

The other important observation is that there have been some alternative self-calibration methods based on the study of stellar fields, which have been successfully implemented in a number of space coronagraphs. Stellar drift scans, in which stars transiting across the field of view during spacecraft motion are used to impose flat-field constraints at the pixel level, have been used on SOHO/LASCO, as described by Llebaria et al.\ (2006) and Gard\`es et al.\ (2013)~\cite{ref-16,ref-21}. An equivalent method using HI (Heliospheric Imager) stellar field observations was shown to work with STEREO by Brown et al.\ (2009)~\cite{ref-30}. The stellar-based techniques are especially strong since stars give well-characterized point sources and their known brightnesses can be used to place restrictions on the flat-field response, and absolute photometric calibration at the same time. The PUNCH mission (DeForest et al., 2026) has taken an even more systematic approach to the problem, incorporating continuous stellar background measurements with a continuous spacecraft rotation to provide ongoing flat-field calibration during regular science operations, demonstrating that star-based flat-field techniques can be integrated into routine mission operations.

\subsection{Comprehensive Comparison of Spatial Flat Field Methods}

Regarding the absolute accuracy, if the inner light source and diffusion plate have been well-designed and aging has been controlled effectively, the error of pixel response and relative flat field calibration often reaches its peak. In these circumstances, it might be appropriate to consider this as a hard reference for the in-orbit flat-field calibration~\cite{ref-2,ref-13}. The attitude roll and bias measurements are the most efficient in defining in-situ vignetting and stray light effects as well as constructing viable models of the overall system response characteristic. The strength of multi-temporal self-consistency methods is that they can be updated efficiently with relatively few additional hardware resources required~\cite{ref-19,ref-25}.

Accurate and traceable absolute flat-field and background models are employed to support the scientific instrumentation of coronagraphy based on the accuracy of the physical parameters inversion such as measuring the coronal heating and recovery of the magnetic field. It is therefore beneficial to combine the attitude-geometry self-calibration method with statistically-based inversion methods, integrating the strengths of internal illumination, attitude-geometry, and statistical inversion approaches~\cite{ref-18}.

Integrating spheres, opal glass, and natural sky techniques are the most common flat-field methods applied during the manufacturing and ground commissioning stages, and they enable systematic decomposition of error sources. The methods using space in the context of flat field have the concept of monitoring and correction in orbit as their main idea. The basis of both of them is similar conceptual framework: The laboratory integrating sphere may be perceived as a model of the ground-based internal light source. The opal glass and natural sky methods can be regarded as ground-based analogues of diffuser plate and offset measurement methods, respectively. Long-term (ground-based/space-based) observations can be regarded as essentially statistical, leveraging software algorithms that distinguish between instrument variations and variations specific to the celestial bodies~\cite{ref-31}. Namely, in space, it is possible to use controlled and actuator-driven structured light fields rather than passive or static fields, which also allows extracting more detailed flat-field information with fewer calibration cycles.

\section{Flat-Field Measurement Method Based on Structured Light Fields via nanoscale Lithography}

\subsection{Nanoscale Lithography Technology and Principles of Structured Light Field Modulation}

The application of nanolithography technologies allows producing periodic or aperiodic structures with a few-nanometer feature size and less than one-micron structures on transparent substrates (quartz, fused quartz) which can control the amplitude and phase of the incident light beams with precision~\cite{ref-9,ref-10,ref-32,ref-33}. First, level the surface and then spin-coat a layer of photoresist. Subsequently, a master mask is produced by electron-beam lithography or laser direct writing, and then a photoresist microstructure is formed. Reactive Ion Etching (RIE) or Ion Beam Etching can be used to transfer the pattern to the substrate; then, photoresist stripping and passivation are carried out~\cite{ref-12,ref-32}. Newer exposure methods, such as e-beam lithography and maskless lithography, can produce sub-micrometer high-aspect-ratio microstructures on quartz-based substrates and are thus used to fabricate special masks for coronagraphs~\cite{ref-34,ref-35}.

\begin{figure}[H]

\includegraphics[width=\textwidth, keepaspectratio]{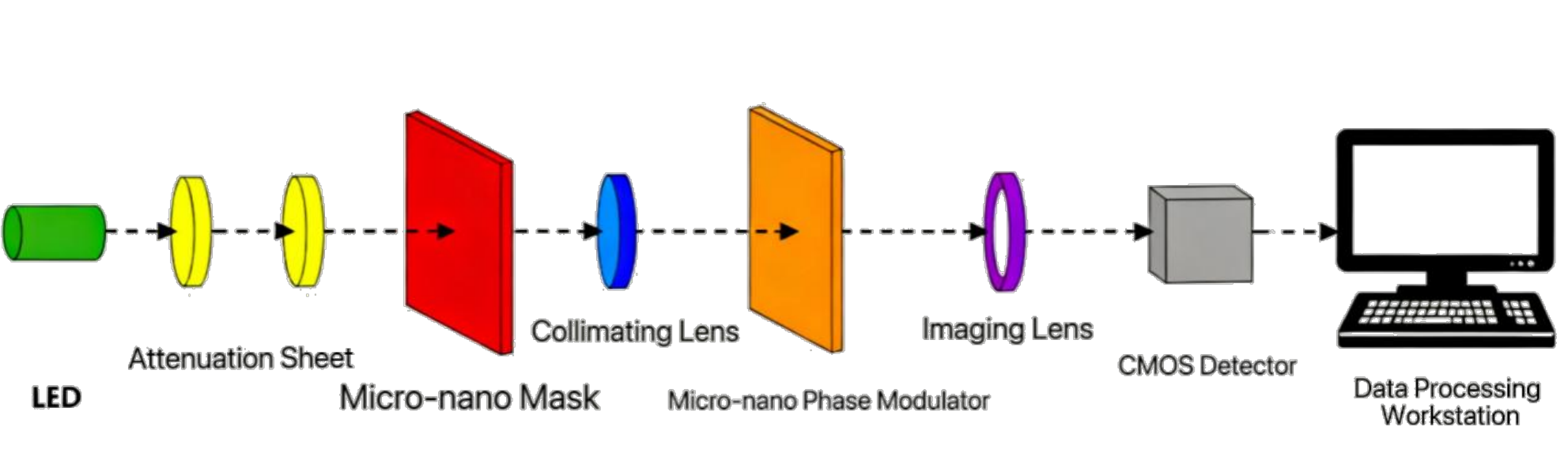}
\caption{Schematic of mask transmittance and phase modulation based on nanoscale-lithography structured light fields.\label{fig9}}
\end{figure}

The two main ways to create a structured light field using nanoscale structures for light field modulation are: first, transmittance modulation; by changing the aspect ratio of a unit cell, the metal-to-dielectric fill ratio, or both together, the local intensity of the transmitted light is altered to produce a spatially coherent intensity modulation profile. The second is phase modulation; that is to say, by adding components with different thicknesses and refractive indices at various positions, the optical path length is modified to control the phase of the wavefront precisely. When the size of the microstructure is smaller than the wavelength of the incident radiation, it can be treated as an effective medium with a continuous change in the distribution of equivalent refractive index. Thus, sub-wavelength anti-reflection coatings, phase plates and metasurface elements can be realised~\cite{ref-36,ref-37}.The resulting structured illumination has high contrast and contains a rich set of high-spatial-frequency components at the focal plane. Although the older method of sphere or diffuser plate integration can achieve a relatively even light distribution to some extent, structured light fields can also vary the spatial distribution of spectral characteristics and contrast. This is a two-edged sword; on the one hand, some spatial frequency components help improve the system's modulation transfer function (MTF) in a particular frequency range, so now a flat-field measurement can achieve pixel-level resolution of response and optical errors. Nevertheless, the differences in intensity at various places set stronger limitations on the inversion model. It enables estimating the pixel response functions very accurately with only a few exposures, which greatly enhances the accuracy of the flat-field calibration~\cite{ref-11,ref-38,ref-39,ref-40}. The schematic of the mask transmittance and phase modulation principles based on nanoscale-lithography structured light fields is illustrated in Figure~\ref{fig9}.
\unskip

\subsection{Theoretical Model for Flat-Field Measurement of Structured Light Fields in a Coronagraph}

The coronagraph in the structured-light-field method can be modelled as a serial linear system of a mask, an optical system and a detector. The incident light field passes through the nanoscale mask and is produced as follows~\cite{ref-13,ref-41}.
\begin{equation}
U_m(x,y) = U_i(x,y)\,T(x,y) ,
\end{equation}

In particular, the system point spread function $h(x,y)$ is calculated based on the mask transmission function $T(x,y)$ as follows. Firstly, the complex amplitude at the pupil plane is modulated by $T(x,y)$ that contains both amplitude and phase information defined by the nanoscale mask design. Then the system PSF is calculated as the squared modulus of the Fourier transform of the pupil function $P(u,v)$ modulated by $T(u,v)$:
\begin{equation}
h(x,y) = \bigl| \mathcal{F}\,P(u,v)\,·T(u,v)\, \bigr|^2
\end{equation}

In which $P(u,v)$ is the pupil aperture function including the optical aberrations (spherical aberration and coma), $\mathcal{F}\{\cdot\}$ is the Fourier transform operator, and $(u,v)$ are the pupil plane coordinates. This formulation, based on standard Fourier optics theory~\cite{ref-6}, ensures that $h(x,y)$ correctly reproduces both the aperture diffraction and the wavefront modulation introduced by the nanoscale mask.
\begin{equation}
S(x,y) = \eta(x,y)\,\bigl|U_0(x,y)\bigr|^2 + N(x,y) 
\end{equation}

The term $N(x,y)$ represents the sum of readout noise and thermal noise, which can be modeled as Gaussian noise with near-zero mean.

The goal of flat-field measurement is to estimate the pixel response function and the nominal point spread function of the system simultaneously. Once these parameters have been defined and spatially localized, it becomes possible to invert the system's nominal point spread function from a small number of structured illumination measurements. A further simplification of the convolution calculation can be obtained by applying Fourier optics, which transforms spatial convolution into multiplication in the frequency domain:
\begin{equation}
\mathcal{F}\{S_k\}(v_x,v_y) \approx \mathcal{F}\{\eta\} \times  \mathcal{F}\bigl\{|U_{ok}|^2\bigr\} 
\end{equation}

Here $H$ is the optical transfer function of the system. The pixel response function $R(x,y)$ is expanded in basis functions or, in some cases, piecewise constants, and the resulting problem can be formulated as a linear or weakly-nonlinear least-squares problem~\cite{ref-42}. The simultaneous solution over multiple different structured-light-pattern sets can provide an estimate of $R(x,y)$. The machine learning regression model used in this paper is a lightweight U-Net convolutional encoder-decoder network. U-Net architecture is a combination of a contracting encoder path and a symmetric expanding decoder path with skip connections allowing the network to encode both overall low-frequency flat-field gradients and local pixel-level response non-uniformities at once. The physics-based Fourier optics flat-field estimate is used as the input to the network to constrain the learning space and ensure physical interpretability. The network is trained to predict the residual term $\epsilon(x,y)$, that is, the difference between the analytically estimated flat field and the measured detector output, with a mean squared error (MSE) loss function. Train on a set of structured illumination exposures taken with different mask patterns and add more data augmentation to help it generalize better. The above two-pronged strategy can help the final flat-field estimate be both physically accurate according to the Fourier optics model and flexible enough to capture any residual sub-pixel non-uniformities not covered by the analytical model through deep learning~\cite{ref-11,ref-12}.

By restricting the learning space of the network and using convolutional neural networks, such as lightweight U-Net architectures, to obtain the residual between the calculated and measured flat fields, this way can improve the accuracy of inversion without sacrificing physical interpretability.

\subsection{Experimental System Design and Key Parameters}

\begin{figure}[H]
\includegraphics[width=\textwidth, keepaspectratio]{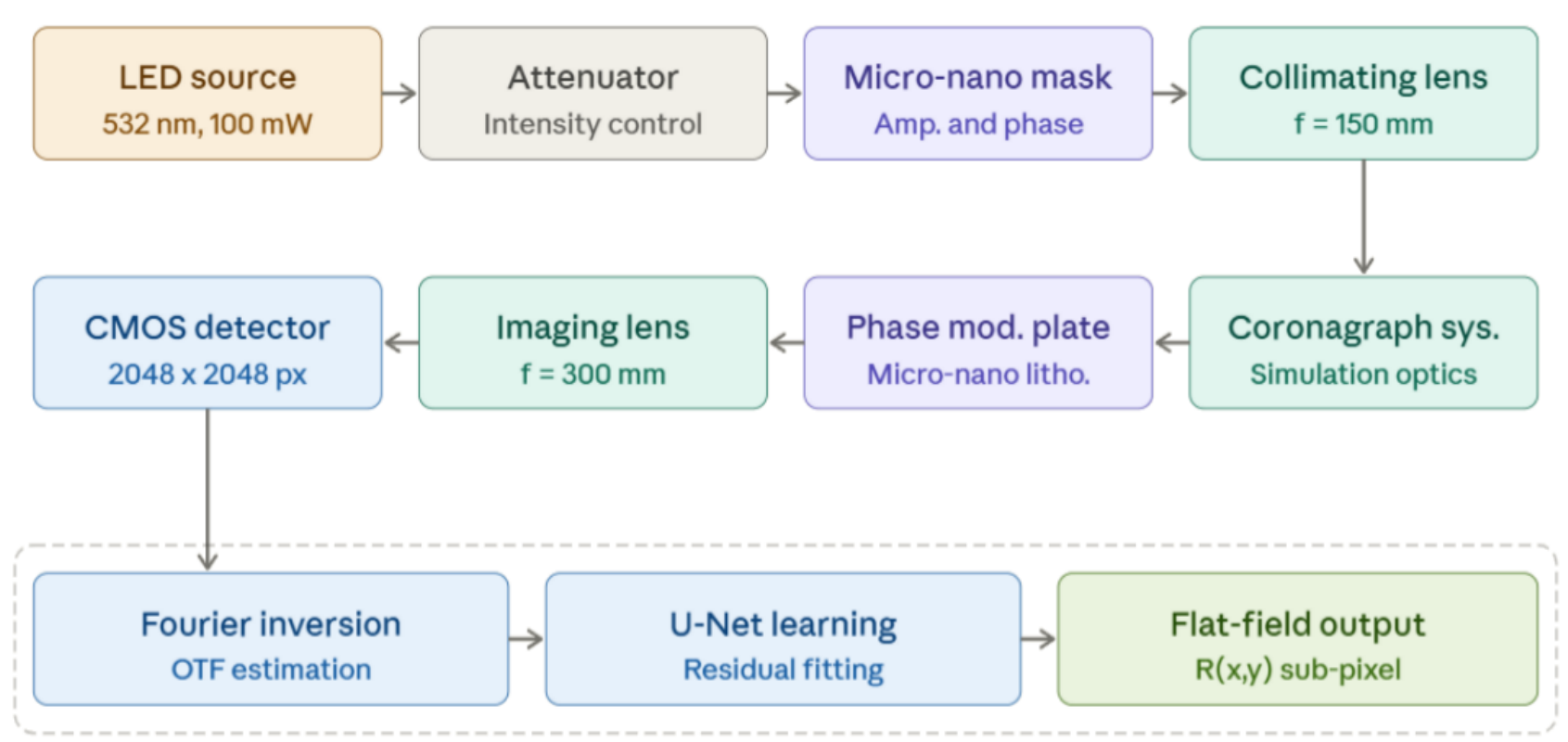}
\caption{Schematic diagram of the experimental optical system for structured light field flat-field measurement of a coronagrap.\label{fig10}}
\end{figure}

The experimental system consists of a module containing the light source, a nanoscale mask, a coronagraphic simulation optical system, a detector and a data processing unit. 
The full optical path configuration is: light-emitting diode (LED) light source $\rightarrow$ attenuator $\rightarrow$ nanoscale mask $\rightarrow$ collimator $\rightarrow$ nanoscale phase modulation plate $\rightarrow$ imaging lens $\rightarrow$ complementary metal-oxide-semiconductor (CMOS) detector, as shown in Figure~\ref{fig10}.
\unskip

The light source is highly uniform at a center wavelength of 532~nm with an intensity of around 100~mW, and its field uniformity exceeds 99.5\%. This arrangement simulates quasi monochromatic irradiation within the range of wavelengths where the coronal measurements took place~\cite{ref-10,ref-12}. The carrier substrate is quartz with a thickness of 1~mm. It is fabricated by electron-beam lithography and reactive ion etching, with a feature size of 500~nm and a transition slope greater than 0.3~$\mu$m.

A collimator with a length of 150~mm and an imaging lens having a focal length of 300~mm and a numerical aperture of about 0.15 are used. A phase-modulation plate with nanoscale lithography is used as the imaging lens and functions as the real aberration-correction component placed inside a coronagraph. 
The detector is 2048$\times$2048 pixels, and the size of one pixel is 3.75~$\mu$m~$\times$~3.75~$\mu$m. Quantum efficiency $\geq 90\%$, read-out noise $\leq 2$~e$^-$, frame rate 30~fps, and thus meets the requirements for both multi-frame fast acquisition and multi-exposure calibration~\cite{ref-2,ref-19}.

The structured light field pattern design consists of:
\begin{itemize}
  \item One-dimensional and two-dimensional periodic gratings are used to probe the response of the modulation transfer function (MTF) of imaging systems at low, medium, and high spatial frequencies.
  \item Radial gradient transfer distributions may be applied to produce a gentle variation in the brightness with respect to heliocentric distance, which renders them more prone to vignetting and unevenness of pixel response.
  \item Multi-frequency checkerboard and ring patterns at three levels incorporate multi-scale spatial properties into a single exposure, improving the conditioning of the inverse problem. With the exchange of portions of masks or plates, one can acquire a series of ordered light that has spatial spectral features that are mutually complementary and have enough useful information to enable further composite inversion.
\end{itemize}

\subsection{Performance Analysis of Flat Field Calibration Effectiveness}

\begin{figure}[H]
\includegraphics[width=\textwidth, keepaspectratio]{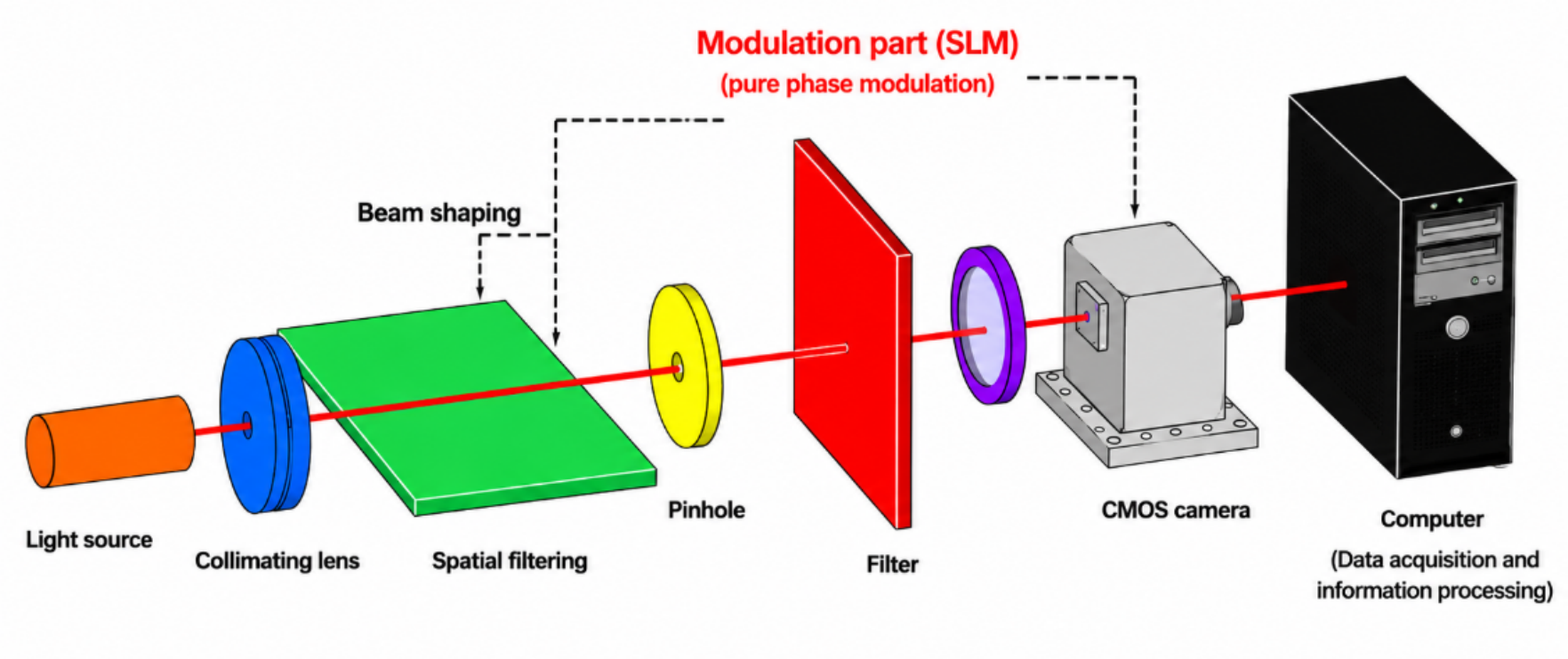}
\caption{Comparison of Flat Field Error Distribution Between Structured Light Field Method and Integrating Sphere Method.\label{fig11}}
\end{figure}

To conduct a flat-field experiment on equal Complementary Metal-Oxide Semiconductor (CMOS) detectors with the same apparatus, three methods were employed; traditional method of even illumination using an integrating sphere, application of frosted glass diffuser plate, and the method of nanoscale structured light field which will be described in the present paper. A central 512$\times$512 pixel region was selected as the region of interest (ROI). The analysis was done in order to get statistics about comparing the error distribution of the three flat-field patterns with the wanted uniform response. Based on the results of the experiments, normalized DN errors, maximum relative errors, and root-mean-square errors of the integrating sphere method can typically be similar to the normative errors found in the standard coronagraph test~\cite{ref-2,ref-7}. the opal glass approach could have slight benefits in repeatability and usefulness in the field. The structured light field reduced the maximum error by an order of magnitude compared with the conventional integrating sphere method, brought the RMS error down to the sub-percent level, and substantially reduced pixel-level variance. As shown in Figure~\ref{fig11}, integrating sphere method demonstrates large regional grouping of flat-field errors, and the structured light field method has higher uniformity of errors with smaller magnitude.

\unskip

Based on the spatial distribution of errors, the residual errors resulting from either the integrating sphere or frosted glass methods exhibit predominantly large-scale gradients or block-like structures ~\cite{ref-43,ref-44}. As shown in the table above, the first few are particularly sensitive to problems such as vignetting of the optical system and localized contamination. On the other hand, the structured light field method has a larger error distribution due to high-frequency random noise, and most large-scale structures are systematically suppressed. Because the structured light fields have many spatial frequencies (which makes them less prone to misidentifying slowly varying aberrations as flat-field terms), the low- and mid-frequency elements of the $\eta(x,y)$ could be restored with significantly greater precision~\cite{ref-6,ref-12}.

\begin{figure}[H]
\includegraphics[width=\textwidth, keepaspectratio]{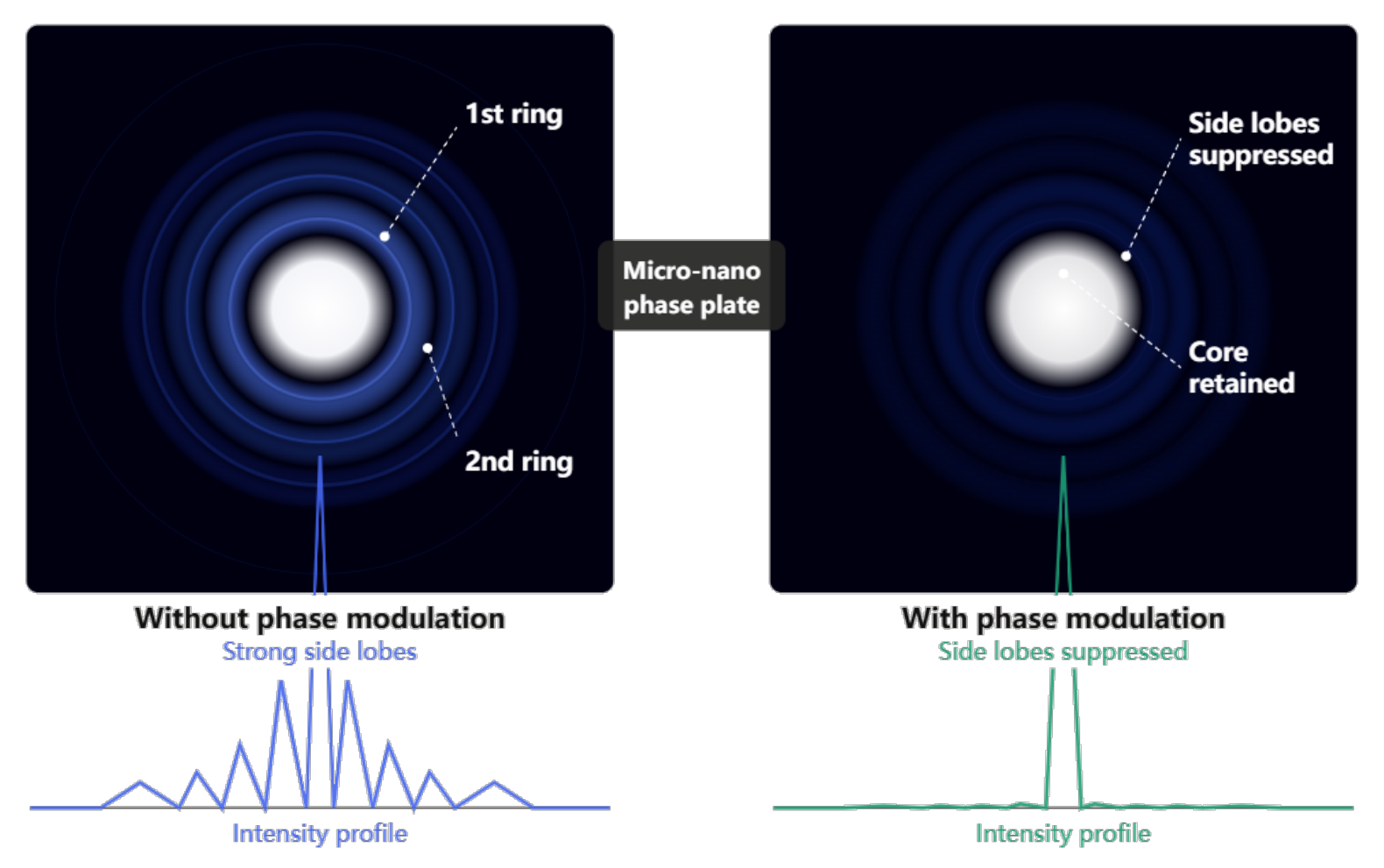}
\caption{Comparison of Airy-spot point spread function (PSF) before and after introducing the nanoscale phase modulation plate, showing central core preservation and side lobe suppression.\label{fig12}}
\end{figure}
\unskip

A nanoscale phase-modulation plate is introduced (as shown in Figure~\ref{fig12}) to reduce the side-lobe intensity of the Airy disk significantly and maintain the central core; thus, a better point spread function is achieved. It is particularly convenient for some cases, such as handling faint coronal features. Additionally, the selection of optimal structured-light-field data, combined with a nanoscale phase modulation plate and suppression of spherical aberration, may be employed to generate higher-quality flat-field models. The parameters of aberration associated with the Point Spread Function (PSF) may then be taken into account, and they will give information about the performance of the optical system under ambient conditions~\cite{ref-13,ref-14}. Stray light and vignetting have been found to be more discriminative with structured light fields and have been improved upon with their application to experiments that mimic illumination at different heliocentric distances, and the work has been done to build an integrated system response model that relies on the concept of a flat field, aberrations, and stray light~\cite{ref-45}.

\subsection{Methodological Advantages and Practical Limitations}

Structured light field measurements of flat fields based on micro-/nano-lithography have been used to improve the accuracy and spatial resolution of calibration. However, with structured illumination, one or more frequency bands of the system can be excited, and since the modulation transfer function (MTF) acts across multiple exposures, it can be used to reconstruct the pixel response function $R(x,y)$. Sub-pixel level non-uniformity is also efficiently measured. In addition, it combines the two types of complementary factors---hardware-based physical priors and algorithmic inversion---and thus avoids the so-called flat-field--scene overlap problem of purely statistical self-consistency algorithms. They can also model aberrations and stray light to provide additional functions that are not offered by the existing integrating spheres and glass diffusers in the initial laboratory calibration. The responses of high-order systems and their discussions in the context of coronagraphs~\cite{ref-4,ref-11,ref-12,ref-46}.

Nevertheless, this solution has a number of engineering constraints. First, the production of high-precision nanoscale masks is relatively expensive and needs to be carried out in a clean room. Each wavelength band may need a different mask or a multi-layer system, and thus fabrication is more complex. Second, it should be accurately placed to achieve a high-precision coaxial alignment with the optical system and remain mechanically stable over time. Thermal expansion and contraction, vibration and contamination can alter the structure of the organised light field pattern~\cite{ref-10,ref-32}. Apply the above method to the in-orbit calibration and ensure that the long-term stability of the nanoscale structures under exposure to space radiation, atomic oxygen and particle deposition is verified. In addition, the relatively small volume and low power consumption of the optical-path design for a switchable structured-illumination device also present extra problems in structural and thermal-control design~\cite{ref-33,ref-46}.

The flat-field measurement method based on nanoscale-patterned structured light fields has proven especially useful as a key technique for high-precision calibration in both laboratory and ground-based stations. It defines the basic flat field model and the system response template of the coronagraph. The addition of internal light sources and the estimation of attitude roll and multipole self-consistent reversal into space applications as a multi-layered calibration system comprising a high-precision laboratory calibration and active in-orbit adjustment. This is why the proposed high-resolution nanoscale structured light field method occupies a key position within the overall flat-field measurement framework rather than serving as an independent component. Such form of integration enables utilizing the advantages of accuracy and physical interpretability and achieving certain equilibrium between cost and engineering possibilities.

It is crucial to define the scope of the structured light field flat-field measurement approach that is suggested in this paper. This approach can be used mainly as a precise laboratory calibration device which is implemented in the pre-launch stage to obtain an exact baseline flat-field model and template of system responses of the coronagraph. It serves this purpose by complementing and enhancing some of the currently available laboratory techniques including integrating sphere and opal glass method, with increased coverage of spatial frequencies and sub-pixel-level characterization of pixel responses.

Concerning its possible use in space, the micro-nano mask assembly might be theoretically installed as an internal calibration unit of a space coronograph apparatus, similar to the internal flat-field lamp and diffuser plate systems now being flown on missions like SOHO/LASCO and Solar Orbiter/Metis. The structured illumination patterns would permit a finer in orbit characterization of detector response nonuniformities than can be achieved by using conventional uniform internal light sources.

The structured light field approach is thus not meant to be used as an alternative to traditional in-flight calibration methods like attitude roll maneuvers, star drift scans, or multi-temporal self-consistency inversion. Instead, it is supposed to form a high-precision base layer of hierarchical multi-method calibration scheme with its physically tightly constrained base, which contributes to the higher accuracy and interpretability of all the following in-orbit calibration improvements.

\section{Comprehensive Comparison and Development Trends of Flat-Field Measurement Methods for Various Types of Coronagraphs}

On the basis of quantitative comparisons and engineering analyses of different approaches to the flat-field problem presented in the preceding sections, it has been found that the future development of the coronagraph flat-field calibration requires a model of calibration, based on a system, that is based on highly accurate baselines and multi-method changes in collaboration.

\subsection{Summary and Categorized Comparison of Advantages and Disadvantages of Primary Methods}

It might be possible that the types of measurement methods in the implementation environment of the coronagraphic flat-field measurements can be divided under the broad category of those adopted in accordance with the manner of implementation, i.e., the ground-based measurement methods and its applications in orbit as space-based measurement methods.

Combining the integration sphere uniform light source method in laboratory experiments provides an extremely uniform and controlled light field that is employed. This suggested approach could also be applied as a preliminary calibration step of an optical system or testing of detectors at the factory. It has been shown to be a highly accurate measurement with good repeatability and it would be akin to the factory reference measurement employed by other coronagraphs and solar telescopes~\cite{ref-47}. But it is big and costly, hence the deployment of it in situ when conducting measurements within a particular time span may be awkward. Or it cannot model the scattering effect of the atmosphere and thermal drift effect measured during the real-life observations as well.

The construction of the frosted glass/diffuser plate process is less complicated and economical. Using installation, commissioning, and regular maintenance of any ground-based coronagraphs it is easy to understand that any territory is flat at high speed. The method involves making measurements according to real optical path partially reflecting optical contamination and optical vignetting of optical system as~\cite{ref-1}. Nevertheless, it also has some disadvantages because it cannot be guaranteed to provide uniformity and repeatability since all of them are influenced by the properties of the diffusers and incidence conditions in light, and the accuracy of the measurement is usually inferior to that of integrating sphere method.

Atmospheric Scattered Light with sunset, twilight, or thin clouds as a Thin Cloud Flat Field System: The suggested approach is based on the scattering of light that travels through air during the sunsets, twilight, or even minimal cloud cover which acts as an extremely rough source of uniform light. It is cheaper to operate with highest possible flexibility of operating stations since it requires small extra hardware. Nevertheless, there is also high sensitivity to weather and atmospheric conditions and therefore becomes highly unstable over time. It is not applicable to precision quantitative flat fielding but can only be used to update trend information and keep track of trends over time.

Scanning Flat Fielding of Solar Disk is a systematic approach to scanning photospheric disks or sun areas in quiet state. Its name suggests that it combines changes in an observation geometry together with previous brightness models to remove contributions made by flat field and background scattering. It works on the actual observation channels and light paths where it can be applied as fine calibration on specific imagers~\cite{ref-5,ref-7}. Nevertheless, it uses the assumption that the solar brightness models are right, but it is sensitive to the fluctuations of solar activity in the short term.

Integrated internal light source and diffuser systems for in-orbit flat-field acquisition: solar instrumentation systems such as SOHO/LASCO or SDO/AIA usually include internal flat-field lamps and diffusers to monitor periodic detector gain drift and optical contamination~\cite{ref-48,ref-49}. This way is convenient, and the systems will not be used for astronomical observations. However, there is a limitation in optical-path coverage, and simulating the entire scientific optical path is not feasible. Among all of the above stages in the imaging path, the detector and the front-end optics account for a relatively large proportion of the flat-field correction. In addition, a prolonged reduction in the light-emitting ability of the source and contamination of the diffuser may cause a gradual increase in errors.

The attitude roll or sun offset observation method is executed as follows: the satellite is turned to rotate on its optical axis or move off the solar centre to perform an experiment of which one could describe as the solar disk or corona effect on the focal plane. The resulting output is integrated with multi-angle information to invert the signal self-consistently~\cite{ref-26}. It performs well in reducing large-scale halo effects and addressing directional scattering when required. However, it consumes attitude control resources, may conflict with observation schedules, and requires a moderate degree of algorithmic complexity.

Multi-temporal self-consistency flat-field analysis: the latest spatio-temporal joint inversion schemes developed for reconstructing the corona and heliosphere rely on statistical algorithms that incorporate multiple time series to isolate more consistent system responses from the significantly varying solar corona images. It is less dependent on the hardware and it needs substantial quantities of data and strict requirements. If aliasing exists between the flat field and structures at high frequencies, the system tends to be biased due to high noise levels~\cite{ref-50}.

Depending on the mechanisms upon which they are based, flat-field methods may be classified into the following types. Hardware-based solutions include integrating spheres, opal glass, in-situ light sources, and nanoscale structured illumination fields. It is based on either physical homogeneity or programmable calibration light sources. They are physically interpretable and can serve as high-precision reference standards in the laboratory. On the other hand, they have the following deficiencies: high cost, difficulties in manufacturing, and limited adaptability for in-field or in-orbit modifications. Solar disk scanning, attitude roll, and off-corona offset observations are all geometry-based observation methods that supplement flat-field measurements by taking advantage of geometric differences in the light source or instrument attitude, and thus automatically account for atmospheric scattering, system vignetting, optical contamination, etc. However, their use depends on time (observation time windows), it depends on the resources of attitudes, and can require more complicated geometric models. Statistical inversion techniques include multi-temporal self-consistency flat fields and joint inversion algorithms. They are based on the principle of using large datasets and statistics to infer system responses, since they extract information from scientific measurements. Low hardware costs also enable long-term in-orbit variation studies~\cite{ref-23,ref-24,ref-47}. However, it is less physically interpretable, as it depends heavily on models, its initial conditions are difficult to control, and it is data-driven.

With respect to the machine learning part stated previously, the estimation of the pixel response function $R(x,y)$ in the structured light field method uses a lightweight U-Net convolutional encoder-decoder architecture~\cite{ref-11,ref-12}. The network utilizes physics-based Fourier optics flat-field estimate as an input and is trained to predict the residual term $\epsilon(x,y)$---which is the difference between the analytically calculated flat field and the real measurement of detector output---with a mean squared error (MSE) loss. The encoder path encodes multi-scale spatial characteristics of the flat-field non-uniformity, whereas the decoder path with skip connections reconstructs the pixel-level response map at full resolution. The two-step design guarantees that the last flat-field estimate is advantageous to both the physical precision of the Fourier optics model and the flexibility of deep learning to capture residual sub-pixel-level non-uniformities not described by the analytical model.

With regard to accuracy, cost and update capability, the overall impression is as follows. In laboratory conditions, there are various methods for generating the baseline flat field. The best practices on how to measure the so-called baseline flat field are with the use of hardware, specifically, an integrating sphere and frosted glass, or with nanoscale structured light fields. On ground station level, frosted glass and natural sky and solar disk scans should be used together as part of a regular routine calibration maintenance at ground station level. In case it is feasible, nanoscale structured light fields are capable of enhancing the accuracy significantly. For space missions, the typical framework combines an internal illumination system with attitude roll or offset measurements, complemented by multi-temporal statistical inversion to achieve in-orbit flat-field correction.

\subsection{Conceptual Framework for a Multi-Method Collaborative Flat-Field Calibration System}

In engineering practice, adopting a single method is unlikely to satisfy the flat-field calibration requirements at all stages of the life cycle of any individual coronagraph during its design and operation. The hierarchical cooperative approach in the calibration plan will be effective and realistic. A system of this kind will have the following modules: high-precision calibration in the laboratory, field updates, and in-orbit self-calibration.

Instrumental calibration of the coronagraph is carried out at the factory acceptance stage on a high-precision full-optical-path flat-field using the three methods listed below: an integrating sphere, a combination of an integrating sphere and opal glass, and a structured light field based on nanoscale lithography.

The integrating sphere can achieve a very high level of uniformity and is thus suitable for normalising the detector gain and the response of the optical system to large-scale signals.

Adjustable spatial frequency fields at the micro-/nano-scale have spectral manipulation capabilities that overcome the limitations of integrating spheres in addressing high-frequency non-uniformity, and can produce flat-field patterns at the pixel level or finer.

They form a combined laboratory flat-field model with aberration and stray light, which can be taken as an anchoring point to the other calibration procedures. Such a model should be regarded as the synergistic integration of the laboratory baseline flat-field with aberration and stray-light models to provide essential support for subsequent calibration. The above anchors can be employed in the upper-level calibration task to enhance both the accuracy and stability, providing a strong foundation for the rest of the research.

After prolonged operation of ground-based coronagraphs, flat fields may undergo significant changes due to variations in atmospheric conditions, optical contamination, and thermal drift. The measures taken at this stage are:

Opal glass plates and diffuser plates combined with natural sky background can be used for high-frequency, low-cost monitoring of flat fields subject to long-term system degradation.If such conditions are met, solar disk scans or quiet-sun field scans can be performed to obtain high-precision data on the current state of the system.Add also the simple flat fields that were measured during the lab. Composite or differential fitting algorithms are then used to remove the effect of in-situ variations, achieving baseline-plus-incremental flat-field correction.

Satellite coronagraphs operate in the most complex environments and cannot be returned to the laboratory.

An internal light source and diffuser plate are used to monitor detector drift and front-end optical performance on a regular basis, while attitude roll and off-corona offset observations provide self-calibration data without interfering with ongoing scientific observations. The statistical inversion schemes with multi-temporal character can be used to address the low-time-scale self-consistency issue of the flat fields and large objects via long-term data integration~\cite{ref-47,ref-50}.

Applying flat-field templates produced by nanoscale-structured light fields allows these templates to serve as a strong prior during in-orbit activities. The issue of flat-fields and structure overlapping might be addressed through the Bayesian or regularization framework, which is considered as a way of constraining the inversion outcomes in statistics. This is one of the most distinctive features of the new methodology compared with traditional ones.

In this hierarchical structure, hardware-based methods provide physical interpretability and traceability of calibration, while algorithmic methods such as statistical inversion, machine learning, and Bayesian updating provide dynamic time-varying compensation. These two strategies will coexist, and neither will replace the other~\cite{ref-38}.

High-precision scientific computations like coronal magnetic field inversion, or weak Coronal Mass Ejection (CME) monitors need to reduce the residuals in the system during the laboratory level and transfer the system complexity to the algorithmic level in order to mitigate the inherent unpredictability of celestial phenomena.

Operational space weather monitoring is feasible because the hardware calibration requirements are relatively lower, and further improvements in algorithmic self-consistency and anomaly detection will enable long-term stability and data homogeneity.

\subsection{Future Research Directions for Coronagraphic Flat-Field Measurements}

The trend of calibration and measurement for next-generation wide-band, high-resolution coronagraphs, including large-aperture telescopes and extreme-ultraviolet/soft X-ray imaging systems, is expected to follow the above trends:

The next generation of coronagraphs are expected to be mostly in the far ultraviolet, optical and near infrared, and will also go into the mid infrared when a resolution close to the limit of diffraction is achieved, and is a more feasible near-term objective. The above reasons put the demands on flat-field calibration: cross-band uniformity and the ability to address high spatial frequencies~\cite{ref-50,ref-51}.

In view of the situations presented above, an entire model of optical-detector response must exist. That is to say, the model should present all bands in a single physical framework rather than treating individual bands or optical paths as separate calibration objects~\cite{ref-2,ref-4}. This integrated plan will regulate how levels of differences in flat fields differ across bands and also improve the resilience of the flat field solutions to weak-signal bands.

Recently, due to the development of nanolithography, some physical opportunities have been realised. Many metasurfaces and structured light fields with specific dispersion and spatial-spectral characteristics can be fabricated to achieve tunable and predictable spatial distributions at different wavelengths, thus providing hardware support for unified flat-field calibration~\cite{ref-9,ref-10}. With this architecture, and on top of it, additional improvements may be made by taking into consideration the application of multi-band data joint fitting and regularization constraints to alleviate the effect of low single-band signal-to-noise ratios on uncertainty and thus improve the stability and power of the resulting flat-field solution~\cite{ref-11}.

A future direction in the advancement of adaptive optics (AO) is dynamic wavefront sensing and machine learning systems, and it may become feasible to evolve the flat-field calibration of a coronagraph from offline calibration toward online detection and real-time compensation:

Large-aperture ground-based coronagraphs can use AO wavefront sensing and field-star observations to measure atmospheric and telescope errors that affect the flat field simultaneously, and then dynamically correct for these errors during imaging~\cite{ref-52}.

A deep learning model can be trained to learn how instrument state changes affect the flat field, and thus a soft-sensing method can be employed for prediction and correction of the field. For long-term data-driven forecasting on space platforms, a practical coupled forecasting system can be developed to combine flat-field, stray light and thermal state data, and predict flat-field variations during the calibration-lamp operation cycle.

Micro-/nano-fabrication processes and new sensors are developing at an increasing rate, and soon a new type of approach to flat-field problems will emerge.

In addition, such fabrication may be widely used in anti-reflection components, stray-light control devices, and even structured-illumination elements, integrating multiple functions--flat-field correction, stray-light suppression and spectral selection---into a single element.

In addition, the new generation of high-quantum-efficiency, low-noise detectors has also increased the uniformity of pixel response and gain consistency. Therefore, the role of flat-field calibration will gradually change from correcting hardware deficiencies to achieving high-precision alignment of the system response and the observed scene model.

Radiation-hardening methods and onboard self-test instruments will also serve as equal enabling factors for the extended flat-field stability of space missions.

Flat-field measurement of coronagraphs is entering a new stage where single-method and local optimisation are no longer the focus, and coordination among multiple methods and system-level optimisation have become the focus. The flat-field measurement method proposed in this paper is based on nanoscale-lithography structured light fields and should be considered a high-precision reference layer in this development. The proposed method does not aim to replace traditional methods that use spheres, opal glass or in-orbit self-consistency. This way will provide a strong foundation and a starting point for the whole flat-field calibration system, increase the resolution of the flat-field models, and add more physical constraints to them.

\section{Conclusion}

Flat-field calibration is needed to obtain accurate quantitative data from coronagraphic observations of the corona.The reliability of the above process depends on its accuracy; therefore, it will affect the reliability of the resulting coronal brightness distribution, coronal mass ejection parameters and magnetic field inversions. The imaging chain of a coronagraph and the sources of flat-field errors in this chain are addressed in this paper. Briefly introduce the basic ideas, construction methods and application scenarios of all kinds of flat-field measurement techniques in ground-based and space-based environments. Each of the above methods has different strengths and weaknesses, and they are not all suitable for all applications; thus, a single method is not ideal.

High-precision baseline calibration is relatively suitable for integrating spheres and laboratory-uniform light sources. Opal glass diffusers and natural-sky backgrounds can be used to support the construction of a ground-based station, as well as other types. Solar disk scanning and field-of-view scanning can generate a near-real observation environment, and both field-of-view response and background can be calibrated simultaneously in very similar operating conditions. For space-based instruments, internal light sources with diffuser plates are used, attitude roll is combined with off-corona offset measurements, and multi-temporal self-consistent statistical flat-field algorithms are employed to update the flat-field in orbit. Nevertheless, they are not without their own deficiencies, such as incomplete coverage of the optical path, restrictions imposed by satellite attitude and observation capabilities, and the accuracy of statistical methods being dependent on assumptions.

Given the above constraints, we propose a laboratory measurement scheme that employs structured light patterns generated by nanoscale-lithography systems, and under controlled structured illumination with high-resolution modulation by micro-/nano-lithography masks, machine learning algorithms are used to obtain the inverse function of the detector response at the pixel or sub-pixel level. The proposed method has a relatively high spatial resolution, is reproducible and physically interpretable. Therefore, it can be used as a flat-field reference in a laboratory to ensure the same initial conditions for ground-based flat-field measurements and to facilitate consistent inter-comparison with space-based measurements. It can also add aberration models and stray-light analysis, and thus be applied to all parts of the imaging path.

\section*{Author Contributions}
\authorcontributions{Conceptualization, Y.F. and X.Z. ; methodology, Y.F., X.Z., and H.L.; software, M.Z.; investigation, Y. L. and T.S.; data curation, M.S. and T.S.; writing---original draft preparation, Y.F. ; writing---review and editing, Y.F. and X.Z.; funding acquisition, X.Z. All authors have read and agreed to the published version of the manuscript.}

\section*{Funding}
\funding{We acknowledge the Chinese Meridian Project for providing high-quality data from the SICG, and the data resources from the National Space Science Data Center, National Science and Technology Infrastructure of 
China. This work was jointly supported by multiple funding sources, the Yunnan Fundamental Research Projects (grant Nos. 202501AS070004 and 202401AT070140), and~the Yunnan Key Laboratory of Solar Physics and Space Science (grant No. 202205AG070009). Specifically, it received partial support from the National Natural Science Foundation of China (NSFC grant Nos. 12173086, 12373063, 11533009, 12163004, 12473089, and~42274227). }

\dataavailability{The data presented in this study are available upon reasonable request from the corresponding authors. The structured light field masks used in this study are not publicly available due to intellectual property protection, but can be provided to qualified researchers under a material transfer agreement.}

\begin{adjustwidth}{-\extralength}{0cm}
\reftitle{References}

\isAPAandChicago{}{%

}

\isChicagoStyle{%

}{}

\isAPAStyle{%

}{}

\PublishersNote{}
\end{adjustwidth}
\end{document}